\documentclass[preprint,1p,times]{elsarticle}

\usepackage{url}
\usepackage[UKenglish]{babel}
\usepackage{amsmath, amssymb}
\usepackage{hyperref}
\usepackage{siunitx}
\DeclareSIUnit\clight{\text{\ensuremath{c}}}
\DeclareSIUnit\clightsq{\text{\ensuremath{c^2}}}
\DeclareSIUnit[per-mode=symbol]\GeVoverc{\GeV\per\clight}
\DeclareSIUnit[per-mode=symbol]\MeVoverc{\MeV\per\clight}
\DeclareSIUnit[per-mode=symbol]\MeVovercsq{\MeV\per\clightsq}
\DeclareSIUnit\bar{bar}
\DeclareSIUnit[per-mode=symbol]\perbar{\per\bar}
\DeclareSIUnit\rad{rad}
\DeclareSIUnit\Sv{\text{Sv}}
\DeclareSIUnit\annum{\text{yr}}
\DeclareSIUnit\CHF{CHF}
\DeclareSIUnit\degree{\text{\textdegree}}

\usepackage{graphicx}

\journal{NIM B}

\begin{document}

\begin{frontmatter}

\title{Particle production and identification for the T10 secondary beamline of the CERN East Area}

\author[a]{M. van Dijk\corref{corr}}
\author[a]{A. Hayat}
\author[a]{D. Banerjee} 
\author[a]{J. Bernhard} 
\author[a,b]{B. Gokturk}
\author[a]{L. Nevay} 
\author[a]{J. Petersen}
\author[a]{M. Schwinzerl}
\cortext[corr]{\href{mailto:maarten.van.dijk@cern.ch}{maarten.van.dijk@cern.ch}}

\address[a]{{European Organization for Nuclear Research (CERN), Esplanade des Particules 1, 1211 Geneva 23, Switzerland. }}
\address[b]{Boğaziçi University, Department of Physics, North Campus, KB Building Floor 3--4, 34342 Bebek/İstanbul, Türkiye.}

\begin{abstract}
  The particle composition of the T10 beam line in the renovated East Hall at CERN has been measured using several experimental techniques and detectors: pressure scans on a threshold Cherenkov counter, a lead-glass calorimeter, time-of-flight, and finally using two separate threshold Cherenkov counters. For the pressure scans, at a given beam momentum, the count rate in the Cherenkov counters is measured as a function of pressure in the counter. The count rate normalized to the rate of beam particles allows computation of the fraction of a specific particle type in the beam. For the method using two threshold Cherenkov counters, one set above and one set below the threshold for the particle species to be identified, with the difference relative to a beam trigger giving the particle fraction for that species. The measurement was proposed in the context of the ``Beamline For Schools'' competition by team Particular Perspective and carried out in 2023. This data was expanded with pressure scans in 2023 and 2025. Overlapping data is compared, leading to a comprehensive overview of the particle content of the T10 beam. 

\end{abstract}

\end{frontmatter}

\section{Introduction to the East Area and T10}

The T10 beam line is a secondary beamline at the CERN East Area. It is driven by \SI{24}{\GeVoverc} protons slow-extracted from the Proton Synchrotron (PS) over a period of \SI{0.4}{\second}, with such cycles flexibly scheduled as needed within the overall PS supercycle. The beam impinges on a so-called multi-target, in which one of five targets can be selected \cite{Bernhard:2792490}. The present experiment was conducted with the beryllium/tungsten target, a \SI{10}{\milli\meter} diameter and \SI{200}{\milli\meter} length beryllium rod, housed in a hollow aluminium cylinder with an inner diameter of \SI{11}{\milli\meter} and an outer diameter of \SI{20}{\milli\meter}. It is capped on the upstream side with a \SI{3}{\milli\meter}-thick aluminium cap and on the downstream side with a \SI{3}{\milli\meter}-thick tungsten cap. This has become the standard target for operation of the T10 beam line since the completion of the East Area Renovation \cite{Bernhard:2792490}. The T10 line consists of a series of magnets transporting secondary particles produced at a vertical production angle of \SI{35}{\milli\rad} to the experimental zone \cite{Gatignon:2730780}.

The goal of this measurement is the detailed characterisation of the particle composition at the end of the secondary beam line over the full momentum range, from \SI{0.5}{\GeVoverc} to \SI{11.5}{\GeVoverc}. Collimators allow for flexible selection of the beam intensity and momentum spread around the central momentum defined by the magnet settings. The collimators allow to select a momentum spread of between \SI{0.6}{\percent} and \SI{15}{\percent} and, depending on the selected momentum, an intensity ranging from some \num{1e3} to close to \num{1e7} per \SI{0.4}{\second} spill. The beam line can transport particles of either positive or negative charge. Neutral particles are rejected due to the non-straight nature of the line but could give rise to additional background. Particles other than electrons, muons, pions, kaons, and protons (as well as their anti-partners) are few and are not considered in this work. 

The T10 beam line is equipped with a set of detectors, including two scintillator counters and two threshold Cherenkov counters, which are designated in the CERN internal naming system as ``XCET''. Material along the beam axis attenuates the beam intensity and deteriorates the beam definition. The material along the beam line sums up to \SI{8.6}{\percent} $X_0$, where $X_0$ denotes the radiation length. The main contributors to this are the length of air downstream the target (\SI{1.7}{\percent} $X_0$), the two thick aluminium windows of the high pressure XCET (\SI{3.2}{\percent} $X_0$ total), and the two beam line scintillator counters (\SI{2.0}{\percent} $X_0$ total). The remaining material (\SI{1.8}{\percent} $X_0$ total) consists of vacuum windows, short lengths of air, and the scintillating fibre detector at the start of the experimental area. The CO$_2$ used in the XCETs would add \SI{0.50}{\percent\per\meter\per\bar}. The gas volume has a length of \SI{3.63}{\meter} for the high pressure XCET (up to \SI{16}{\bar}) and \SI{3.25}{\meter} for the low pressure XCET (up to \SI{4.2}{\bar}). When empty, the pressure is sufficiently low (below \SI{0.05}{\bar}) that the remaining gas adds less than \SI{0.1}{\percent} $X_0$ for either XCET. The maximal added material budget is \SI{29.0}{\percent} $X_0$ for the high pressure XCET and \SI{6.8}{\percent} $X_0$ for the low pressure XCET.

The measurement of the content of the particle beam was proposed by the team ``Particular Perspective'' in the 2023 edition of the Beamline For Schools competition (BL4S \cite{bl4s_homepage}). The fixed-point XCET measurements, the TOF and the lead-glass calorimeter measurements were taken with the student team, and further expanded with the Cherenkov threshold scans.

\section{Setup and Experimental Methods}
\label{sec:expMethodSetup} 

The existing beam instrumentation was used in conjunction with a further set of organic scintillator counters and lead-glass calorimeters to allow particle identification over the full range of momenta. In all cases, the trigger was defined by the coincidence of at least two scintillator counters. 

Each of the four experimental methods used a different subset of detectors: 
\begin{itemize}
    \item The first method uses two XCETs at two different pressures, one above and one below the threshold pressure of a given particle. Recording the hits in a narrow time window around the trigger in the two XCETs and computing the difference between the number of hits then allows for the fraction of a given particle species to be defined. This measurement technique will be referred to as fixed-point XCET.
    \item The second method employs a single XCET, which is scanned in pressure, counting the ratio of the coincidences with a fixed trigger. The levels that can be identified in the resulting pressure scan allow the derivation of particle fractions. These methods fall short for some groups of particles, see also Sec.~\ref{sec:expSetup_T10_XCETs}. This method will be referred to as the XCET scan. 
    \item The third method uses a lead-glass calorimeter for the measurement of the electron/positron fraction. 
    \item The final method is time-of-flight (ToF), used for identification of protons at low energy.
\end{itemize}

In Sec.~\ref{sec:expSetup}, the experimental setup is laid out. In Secs.~\ref{sec:expSetup_T10_XCETs} and \ref{sec:expMethodToF}, the various methods and components will be described after a brief discussion of the evaluation of the refractive index of the gas in the XCET in Sec.~\ref{sec:XCET_refractive_index}.

\subsection{Experimental Setup}
\label{sec:expSetup} 

\begin{figure}[!htb]
    \centering
    \includegraphics[width=0.95\textwidth]{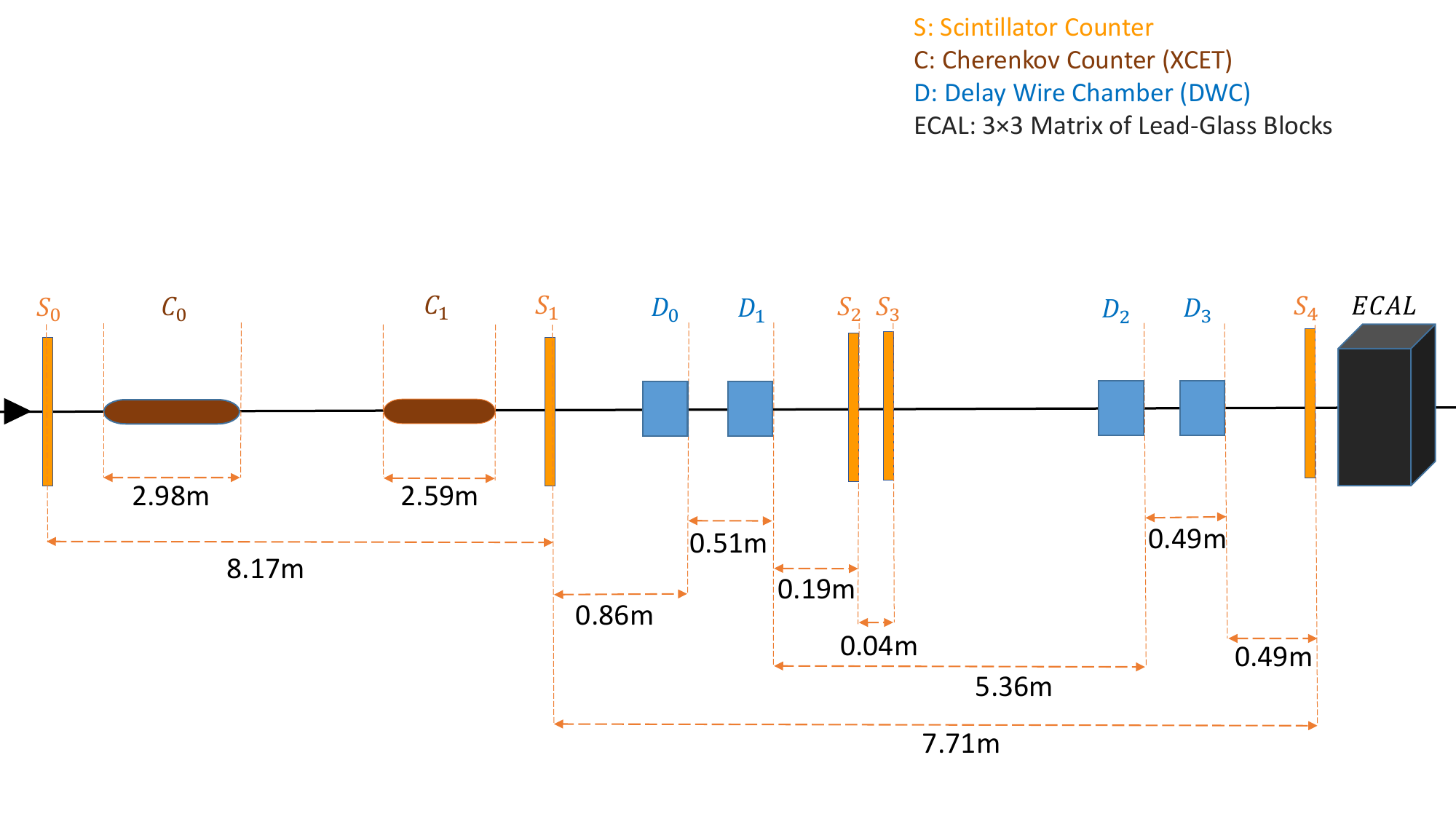}
    \caption{Layout of the BL4S experiment with the distances between all elements, including those used for the XCET pressure scans.} 
    \label{fig:pakistanExperimentalLayout}
\end{figure}

The full setup employed for the experiment consisted of five scintillator counters, two threshold Cherenkov counters, two pairs of Delay Wire Chambers (DWCs), and a 3$\times$3 matrix of lead-glass blocks serving as an electromagnetic calorimeter (ECAL). The setup and the relative positions of the elements are shown in Fig.~\ref{fig:pakistanExperimentalLayout}. 

The two beam line scintillator counters are labelled here as S$_0$ and S$_1$, and the two Cherenkov counters as C$_0$ and C$_1$. The other detectors--- additional scintillator counters, DWCs, and ECALs---were installed specifically for the purposes of this experiment. The signals from the five scintillator counters, S$_0$ to S$_4$, and the Cherenkov counters, C$_0$ and C$_1$, are recorded using a CAEN V1290N Time-to-Digital Converter (TDC), while the analogue signals from the lead-glass calorimeters are recorded using a CAEN V760N Charge-to-Digital Converter (QDC). The S$_2$ and S$_3$ scintillator counters as well as the DWCs were used for beam alignment and not employed directly in the experiment. The TDC is gated by the event trigger, which is the coincidence between the S$_1$ and S$_4$ scintillator counters and records signals during the programmable time window of the TDC. Due to limitations of the data acquisition (DAQ) system, the readout rate is limited to approximately \SI{1.2e3}{\per\second}. Although this is significantly lower than the particle flux, it effectively samples random events throughout the spill. Pressure scans were recorded using an automated pressure scan implemented in the CESAR control software used for operating the beam line \cite{rae:icalepcs2021-tupv047}, detailed further in Sec.~\ref{sec:expSetup_T10_XCETs}. The data is stored on disk and analysed with ROOT \cite{BRUN199781}.

\subsection{Refractive Index of the Cherenkov Radiator}
\label{sec:XCET_refractive_index} 

\begin{figure}[!htb]
    \centering
    \includegraphics[width=0.65\textwidth]{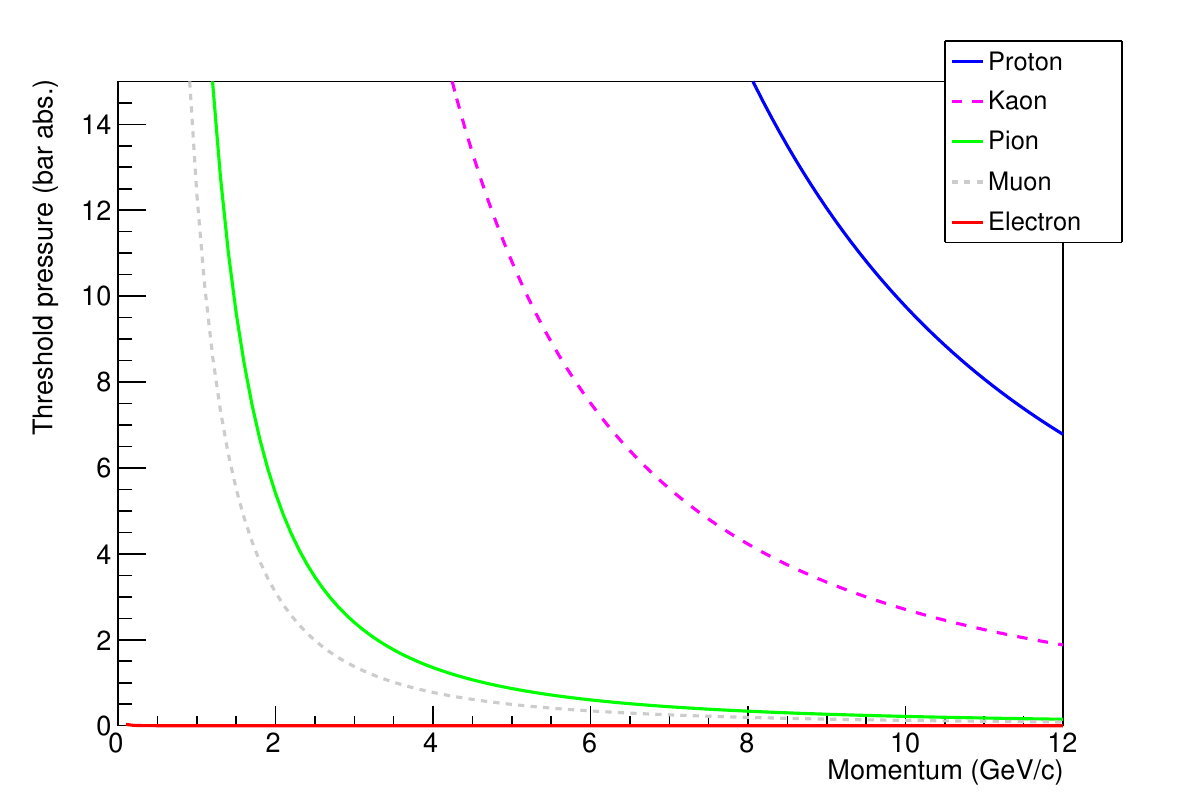}
    \caption{Threshold pressures in CO$_2$ calculated for electrons, muons, pions, kaons and protons using $k$ = \SI{4.5e-4}{\perbar}. } 
    \label{fig:threshold_pressures}
\end{figure}

For the purpose of these measurements the XCETs were filled with $>$\SI{99.9}{\percent} pure CO$_2$. Cherenkov light is emitted under an angle $\theta_c$: 

\begin{equation}
    \mathrm{cos}(\theta_c) = \frac{1}{n \beta}, 
\label{eqn:Cherenkov}
\end{equation}

\noindent
where $n$ is the refractive index of the gas and $\beta$ is the relativistic beta, i.e. the fraction of the velocity of the given particle of the speed of light $c$.  For a gas, the refractive index scales according to the Lorentz-Lorenz equation \cite{CHARITONIDIS201920, CHARITONIDIS2017134}. The refractive index can be approximated as: 

\begin{equation}
    n \approx \sqrt{1+\frac{3AP}{RT}}, 
\label{eqn:Lorentz-Lorenz}
\end{equation}

\noindent
where $A$ is the molar refractivity (constant for a given gas), $P$ is  pressure, $R$ is the gas constant and $T$ the absolute temperature. Equation~\ref{eqn:Cherenkov} gives rise to a threshold phenomenon: Cherenkov light is only emitted when $n\beta > 1$. When increasing the pressure of the gas in the XCET the refractive index will change, and a given particle species at a given momentum will start generating Cherenkov light when the threshold condition is met. Temperature fluctuations in the gas give rise to fluctuations of the refractive index. A change of  \SI{5}{\celsius} would imply a shift of the threshold pressure of up to some tens of millibars and is small compared to the pressure range accessible for the XCET. As the thresholds are not explicitly evaluated from data but only used to calculate indicative pressure settings, temperature fluctuations are not accounted for further. Expanding Eqn.~\ref{eqn:Lorentz-Lorenz} as a Taylor series and retaining only the leading term gives $n=1+kP$, retaining only the dependency on pressure. For CO$_2$ at room temperature and pressure (\SI{20}{\celsius}, \SI{1}{\bar}), evaluated at \SI{255}{\nano\meter} the constant $k$ is around \SI{4.5e-4}{\perbar} \cite{BIDEAUMEHU1973432}. At this wavelength, around half of the generated Cherenkov photons would be detectable by the PMT used. This relation allows for easy evaluation of indicative threshold pressures. Figure~\ref{fig:threshold_pressures} shows the thresholds calculated for electrons, muons, pions, kaons, and protons in the pressure range of the high pressure C$_0$ and the momentum range of T10. 

The refractive index of CO$_2$, evaluated at a given temperature and pressure, still depends on the wavelength of the light under consideration and is higher for lower wavelengths. For any given detector, the UV part of the spectrum will pass the pressure threshold before the longer wavelengths. Even for a mono-energetic  particle beam with negligible size and divergence, there will be a turn-on region rather than a sharp step, dictated by the optical dispersion of the gas convoluted with the transmission efficiency of any optics and the detection efficiency as a function of wavelength. The width of the pressure range over which this rise occurs scales linearly with pressure. 

Another effect is introduced in Eqn.~\ref{eqn:Cherenkov} by $\beta$. The momentum spread in the beam gives rise to a spread to this term, which manifests as a further widening of the turn-on regime, on top of chromatic dispersion due to the gas. Nevertheless, since the thresholds are not determined in absolute terms, the key requirement is that the evaluation pressures lie well beyond the turn-on region.

\subsection{The XCETs of the T10 Beam Line}
\label{sec:expSetup_T10_XCETs} 

\begin{figure}[!htb]
    \centering
    \includegraphics[width=0.95\textwidth]{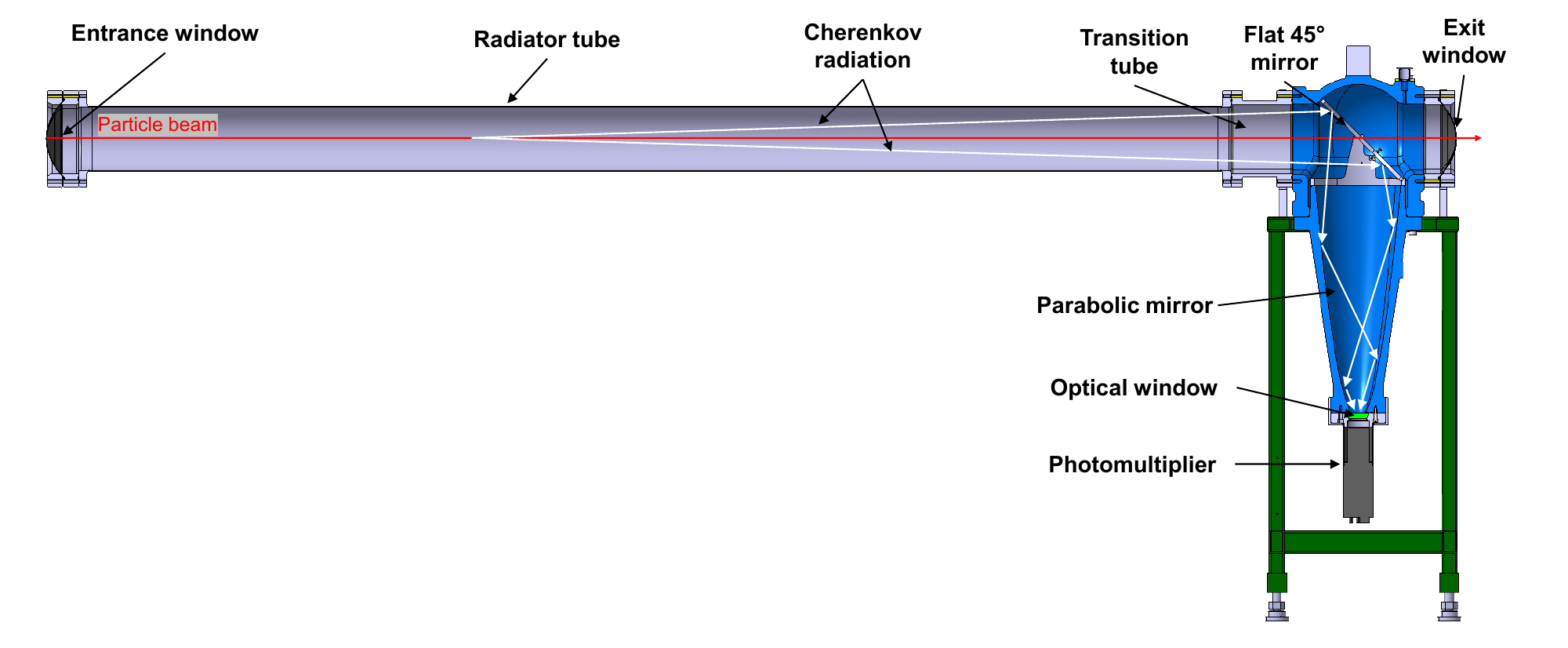}
    \caption{Design of the East Area XCET \cite{buesa_orgaz:ipac2024-wepg90}.} 
    \label{fig:XCET_design}
\end{figure}

The two XCETs of the T10 beam line are both of the same design as shown in Fig.~\ref{fig:XCET_design} and are rated for up to \SI{16}{\bar} (C$_0$) and \SI{4.2}{\bar} (C$_1$), respectively. Measuring over the distance for which the Cherenkov light is collected, C$_0$ has an effective radiator length of \SI{3.34}{\meter}, and C$_1$ an effective radiator length of \SI{2.96}{\meter}.

The radiator tube is filled with gas. The generated Cherenkov light propagates forward and is reflected downward by a flat mirror angled at \SI{45}{\degree}. A parabolic mirror assists with optimal light collection towards the PMT (ET Enterprises 9814QB) placed outside of the gas volume, behind a quartz window. 

For the runs using the XCETs where the pressure was fixed, one XCET was set below the expected threshold (some tens to hundreds of \si{\milli\bar}) and the other XCET to be well above the threshold (half a \si{\bar} or more). A full list of the settings is included in the preliminary analysis \cite{bib:bl4sBeamComposition}. It was observed later in the pressure scans that the pressure thresholds were close to the calculated threshold. Hits are recorded in the two XCETs in coincidence with the trigger, itself being the coincidence of S$_1$ and S$_4$. Computing the difference between number of hits in the XCET above threshold and below threshold then allows for computation of the relevant particle fractions. Errors were considered to come from counting statistics. A global \SI{3}{\percent} error was applied to account for slightly different efficiencies in the XCET detection. 

For the pressure scans, the signal from the XCETs was discriminated by a \SI{-15}{\milli\volt} threshold  in a custom VME board designed for the CERN XCETs \cite{XCET_electronics}. Discrete steps in the detected signal versus pressure graphs were identified and assigned to specific particle species, in accordance with their expected pressure thresholds. Protons at high momenta ($>$\SI{8}{\GeVoverc}) are within the accessible pressure range of the high pressure XCET. At intermediate momenta (between 4.5 and \SI{8}{\GeVoverc}) the proton fraction of the beam is computed as those particles which are not electrons, muons, pions and kaons assuming that the XCET is fully ($>$\SI{98}{\percent}) efficient. Error calculation was based in first instance on the counting statistics and fluctuation across the plateaus, but additional sources of error were found, further discussed in Sec.~\ref{sec:expResults}.

\subsection{Time-of-Flight Measurement and Lead-Glass Calorimeter}
\label{sec:expMethodToF} 

The distance between the scintillation counter S$_0$ and S$_4$ is about \SI{16}{\meter} which allows for a time-of-flight (ToF) measurement of beam particles. This is in particular useful for lower energy protons that are below the threshold of the Cherenkov counters. This method was used to identify the proton fractions in the beam at 1 and \SI{2}{\GeVoverc}. At these momenta, kaons are not considered as they decay in the \SI{45}{\meter} from the target to the start of the experimental zone, around \SI{99.5}{\percent} decaying for \SI{1}{\GeVoverc} kaons and \SI{94}{\percent} for \SI{2}{\GeVoverc} kaons. The TOF difference between electrons and protons at \SI{1}{\GeVoverc} over \SI{16}{\meter} is \SI{19.7}{\nano\second}, and \SI{5.5}{\nano\second} at \SI{2}{\GeVoverc}. The TOF difference between protons and pions is marginally but not significantly different, \SI{19.1}{\nano\second} and \SI{5.4}{\nano\second} respectively. With a measured standard deviation on the TOF distributions of \SI{1.6}{\nano\second}, the peaks are four standard deviations removed from each other and their respective content can be well estimated. 

The electromagnetic calorimeter consists of nine blocks of lead-glass (SF57), each \SI{37}{\centi\meter} long, recovered from the barrel section of the electromagnetic calorimeter of the OPAL experiment at LEP \cite{1991275}. Each lead-glass block has a Hamamatsu R2238 PMT (\SI{76}{\milli\meter} diameter) glued to its end face. The lead-glass represents 24.6 radiation lengths, largely sufficient to contain electromagnetic showers and in particular to identify electrons at energies above \SI{2}{\GeVoverc}. 

The gain of the nine PMTs of the calorimeter was equalized. The total analogue signal from the lead-glass calorimeter is corrected for pedestals and calculated as the sum of the total charges from the nine blocks. The total pulse height is proportional to the energy deposited in the calorimeter. For the identification of electrons and positrons an absolute energy calibration is not required. The electron/positron beam fraction is then calculated by integrating the peak of the pulse height spectrum and dividing by the total number of events. The lead-glass blocks have a nuclear interaction length of \SI{27}{\centi\meter} and therefore a  fraction of the hadrons initiate showers that partially overlap with the electron peak. In the case of overlap, the number of electrons is computed by integrating the right half of the Gaussian shaped electron peak and multiplying by two. The integration method varies as a function of beam momentum. This is taken into account when calculating the systematic errors. More details can be found in \cite{bib:bl4sBeamComposition}.

\section{Experimental Results}
\label{sec:expResults}

The analysis is presented separately for each experimental method. Section~\ref{sec:resultsXCETFixedPlusToF} focuses on the data gathered using fixed pressures in the XCET detectors and the ToF method. Section~\ref{sec:resultsLG} discusses the electron results gathered with the lead-glass calorimeter, and compares these with the data of the preceding section. Section~\ref{sec:resultspressurescan} focuses on XCET scans gathered in 2023, with a complementary set of data gathered in 2025. Following these initial results, the differences in the trigger used for the various methods is compared in Sec.~\ref{sec:trigger_comparison}. Section~\ref{sec:compareAll} contains the final comparison of all data. 

\subsection{Fixed XCET and ToF Results}
\label{sec:resultsXCETFixedPlusToF} 

During the BL4S weeks in September 2023, data on the particle composition was acquired for a range of beam momenta and Cherenkov threshold settings. One XCET was set well above the threshold and the other below the threshold. The results are normalized relative to the trigger, the coincidence of the scintillator counters S$_1$ and S$_4$. The exact pressures used for these runs are not listed here but were included in the initial BL4S data analysis \cite{bib:bl4sBeamComposition}. 

\begin{figure}[!htb]
    \centering
    \includegraphics[width=0.5\textwidth]{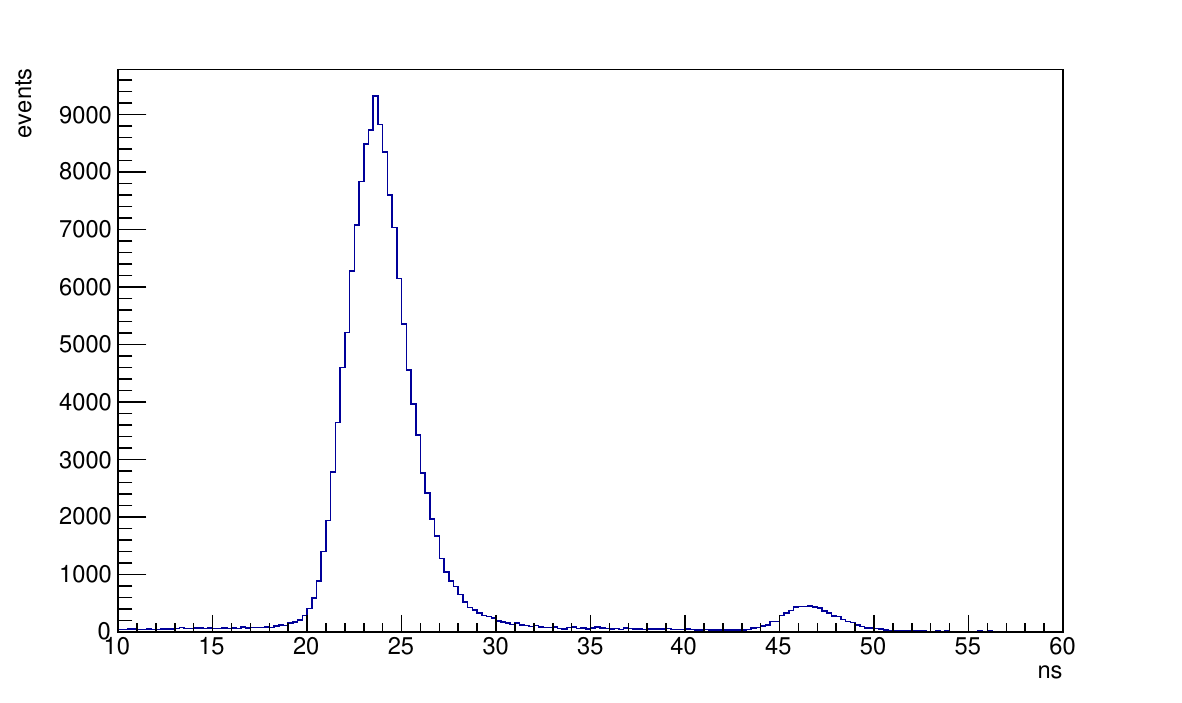}
    \caption{Histogram of TOF between S$_0$ and S$_4$ at \SI{1}{\GeVoverc}. The proton peak is found around \SI{47}{\nano\second}. The peak to the left corresponds to positrons, muons, and pions. At this momentum, kaons are expected to decay before reaching the experimental zone. } 
    \label{fig:ToF1GeVnoElectronsT10}
\end{figure}

An example of proton identification by ToF is shown in Fig.~\ref{fig:ToF1GeVnoElectronsT10}. The proton content is found by integrating the peak around \SI{47}{\nano\second}, assuming no background. It is normalized relative to the number of triggers, which consists of the coincidence of S$_1$ and $S_4$ with S$_0$ added in the analysis, as otherwise the ToF is not defined. The effect of this definition being different to the fixed XCET trigger is expected to be around \SI{1}{\percent} relative to the particle content found. This number is discussed further in Sec.~\ref{sec:trigger_comparison} and added as a systematic error. The ToF difference observed at \SI{1}{\GeVoverc} between the proton peak and the electron/pion/muon peak is compatible with the with the expected flight time over the distance between the scintillator counters S$_0$ and S$_4$, see Fig.~\ref{fig:pakistanExperimentalLayout}. 

\begin{figure}[!htb]
	\centering
	\includegraphics[width=0.49\textwidth]{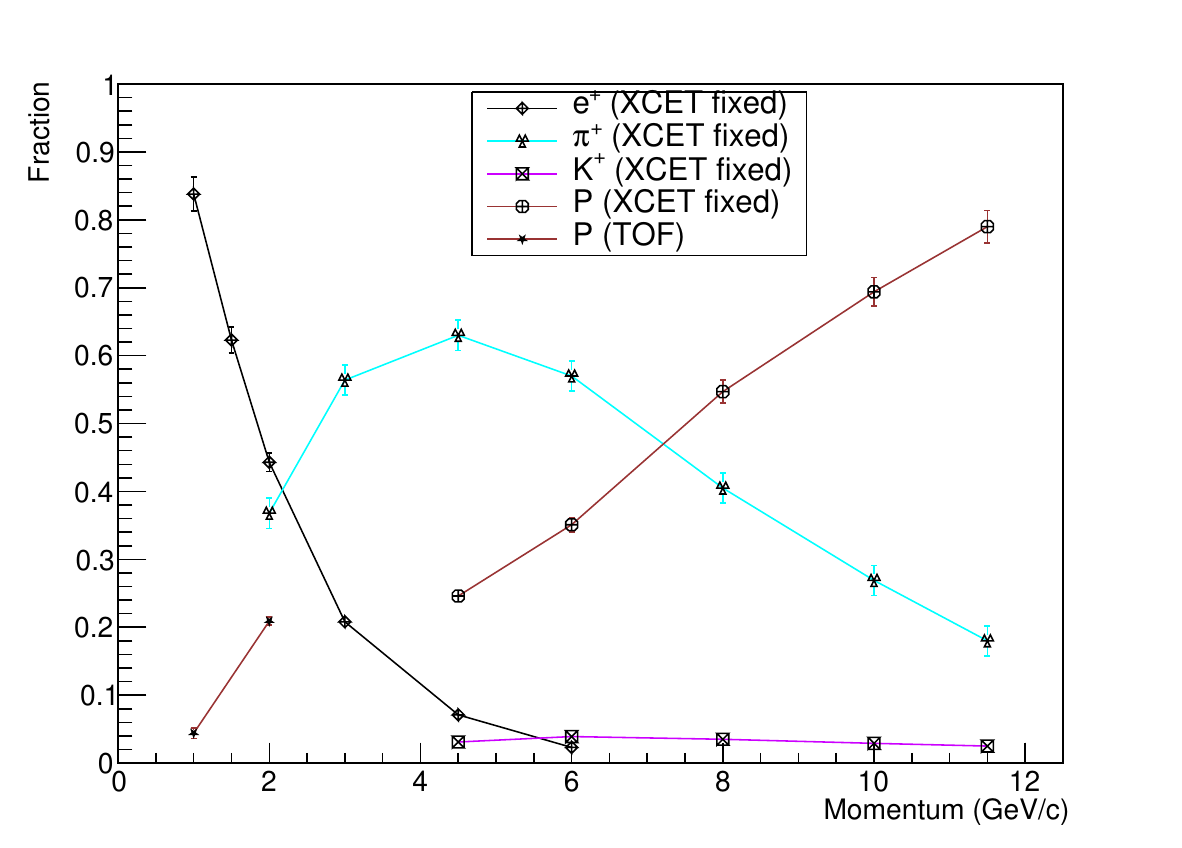}
	\includegraphics[width=0.49\textwidth]{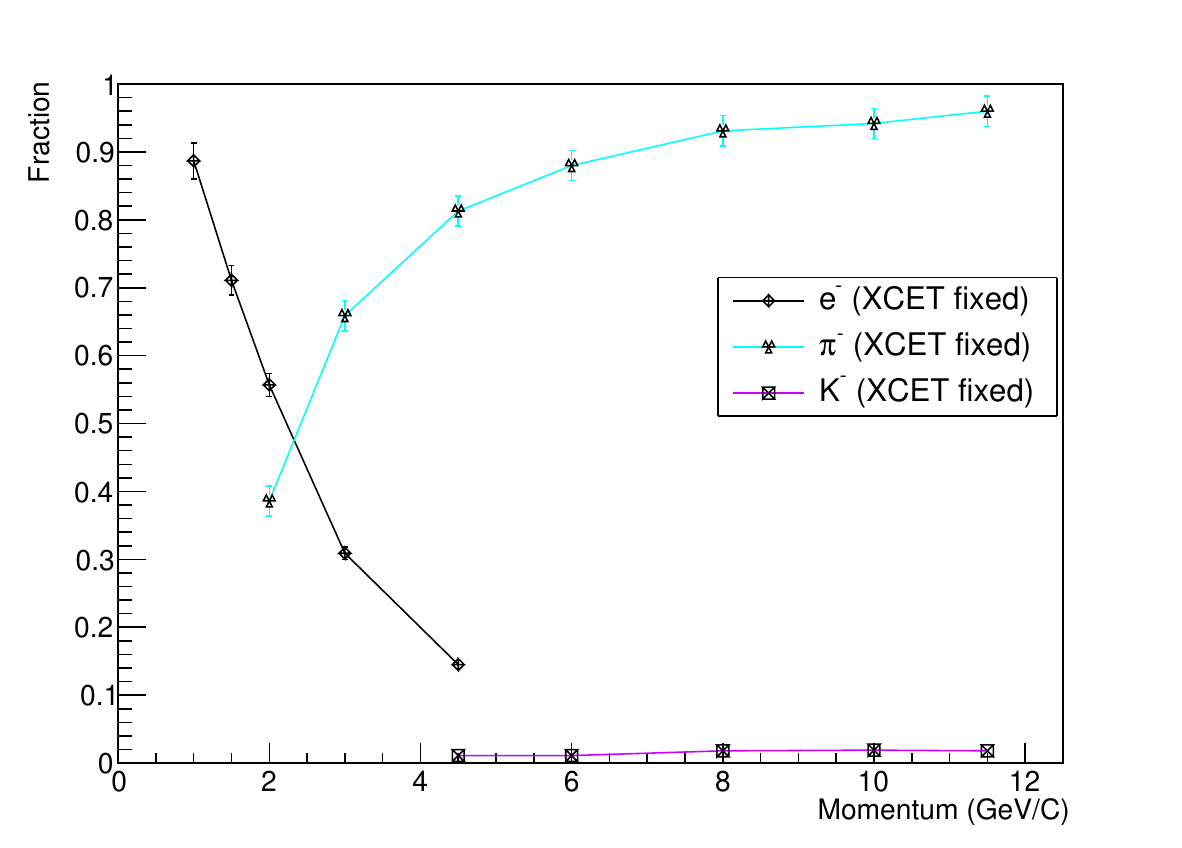}
	\caption{Fractions of positive particles (left) and negative particles (right), XCET fixed. Proton fractions at 1 and 2 GeV/c measured by TOF. Data points joined by straight lines per particle species. } 
	\label{fig:BL4SParticlesInBeamT10}
\end{figure}

The fixed-point XCET and ToF data taken in the 2023 BL4S beam test for all momenta and particles are analysed and shown in Fig.~\ref{fig:BL4SParticlesInBeamT10}. The results are listed in Tabs.~\ref{tab:BL4SresultsPositive} and~\ref{tab:BL4SresultsNegative}.

\begin{table}[htp]
\caption{Fixed-point XCET and ToF results, positive particles. Data denoted with * from ToF}
\centering
\begin{tabular}{c|c|c|c|c}
    p (\unit{\GeVoverc})   & $e^+$                 & $\pi^+$           & $K^+$                 & P                 \\ 
    \hline
    1.0         & 0.838 $\pm$ 0.025    &                    &                       & 0.044 $\pm$ 0.008*\\
    1.5         & 0.623 $\pm$ 0.019    &                    &                       &                   \\
    2.0         & 0.443 $\pm$ 0.013    & 0.368  $\pm$ 0.022 &                       & 0.209 $\pm$ 0.006*\\
    3.0         & 0.208 $\pm$ 0.007    & 0.564  $\pm$ 0.023 &                       &                   \\
    4.5         & 0.071 $\pm$ 0.003    & 0.630  $\pm$ 0.022 & 0.031 $\pm$ 0.004     & 0.246 $\pm$ 0.008 \\
    6.0         & 0.023 $\pm$ 0.001    & 0.570  $\pm$ 0.022 & 0.039 $\pm$ 0.004     & 0.351 $\pm$ 0.011 \\
    8.0         &                      & 0.405  $\pm$ 0.022 & 0.035 $\pm$ 0.003     & 0.547 $\pm$ 0.017 \\
    10.0        &                      & 0.269  $\pm$ 0.022 & 0.029 $\pm$ 0.003     & 0.694 $\pm$ 0.021 \\
    11.5        &                      & 0.180  $\pm$ 0.022 & 0.025 $\pm$ 0.002     & 0.790 $\pm$ 0.024 \\
    \end{tabular}
\label{tab:BL4SresultsPositive}
\end{table}

\begin{table}[htp]
\caption{XCET Fixed results, negative particles}
\centering
\begin{tabular}{c|c|c|c}
    p (\unit{\GeVoverc})   & $e^-$                & $\pi^-$           & $K^-$             \\ \hline
    1.0         & 0.887 $\pm$ 0.027    &                   &                   \\
    1.5         & 0.711 $\pm$ 0.022    &                   &                   \\
    2.0         & 0.557 $\pm$ 0.017    & 0.386 $\pm$ 0.022 &                   \\
    3.0         & 0.309 $\pm$ 0.009    & 0.659 $\pm$ 0.022 &                   \\
    4.5         & 0.145 $\pm$ 0.005    & 0.813 $\pm$ 0.022 & 0.011 $\pm$ 0.005 \\
    6.0         &                      & 0.880 $\pm$ 0.022 & 0.011 $\pm$ 0.005 \\
    8.0         &                      & 0.931 $\pm$ 0.023 & 0.018 $\pm$ 0.005 \\
    10.0        &                      & 0.942 $\pm$ 0.022 & 0.019 $\pm$ 0.004 \\
    11.5        &                      & 0.960 $\pm$ 0.022 & 0.018 $\pm$ 0.005 \\
\end{tabular}
\label{tab:BL4SresultsNegative}
\end{table}

\subsection{Lead-Glass Calorimeter Results and Comparison with Cherenkov Data}
\label{sec:resultsLG}

\begin{figure}[!htb]
    \centering
    \includegraphics[width=0.5\textwidth]{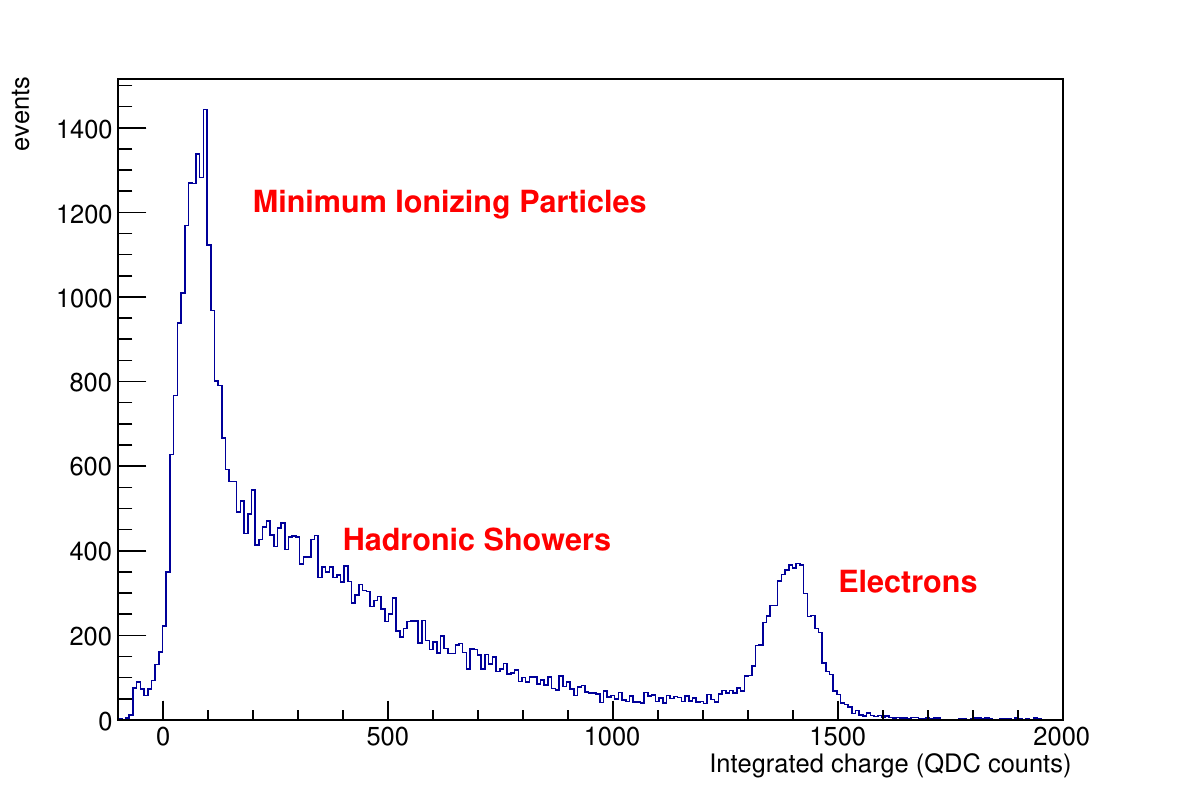}
    \caption{Integrated charge distribution in the lead-glass calorimeter at \SI{4.5}{\GeVoverc}, negative. Sources of the major features of the plot as indicated. }
    \label{fig:LG_total_Integrated_Charge}
\end{figure}

An example of the integrated charge spectrum of the lead-glass calorimeter is shown in Fig.~\ref{fig:LG_total_Integrated_Charge} at \SI{4.5}{\GeVoverc}. Three components can be distinguished. The narrow peak on the left-most side of the distribution consists of muons and hadrons that did not shower (together the minimum ionizing particles). The non-showering hadronic component is substantial as the detector has a thickness of only 1.4 nuclear interaction lengths. The broad component to the right of this peak consists of the hadrons that produced a shower which is not fully contained in the detector. The right-most peak corresponds to the electrons. The electron/positron fraction is calculated by integrating this latter peak. If the peak is well separated from the hadron pre-shower, the integration is done over all the bins in the peak. In case of overlap with the hadron pre-shower component, the central symmetric part of the peak is integrated and then the Gaussian tail at the high momentum side. The latter is multiplied by two and added to the central part.

\begin{figure}[!htb]
    \centering
    \includegraphics[width=0.49\textwidth]{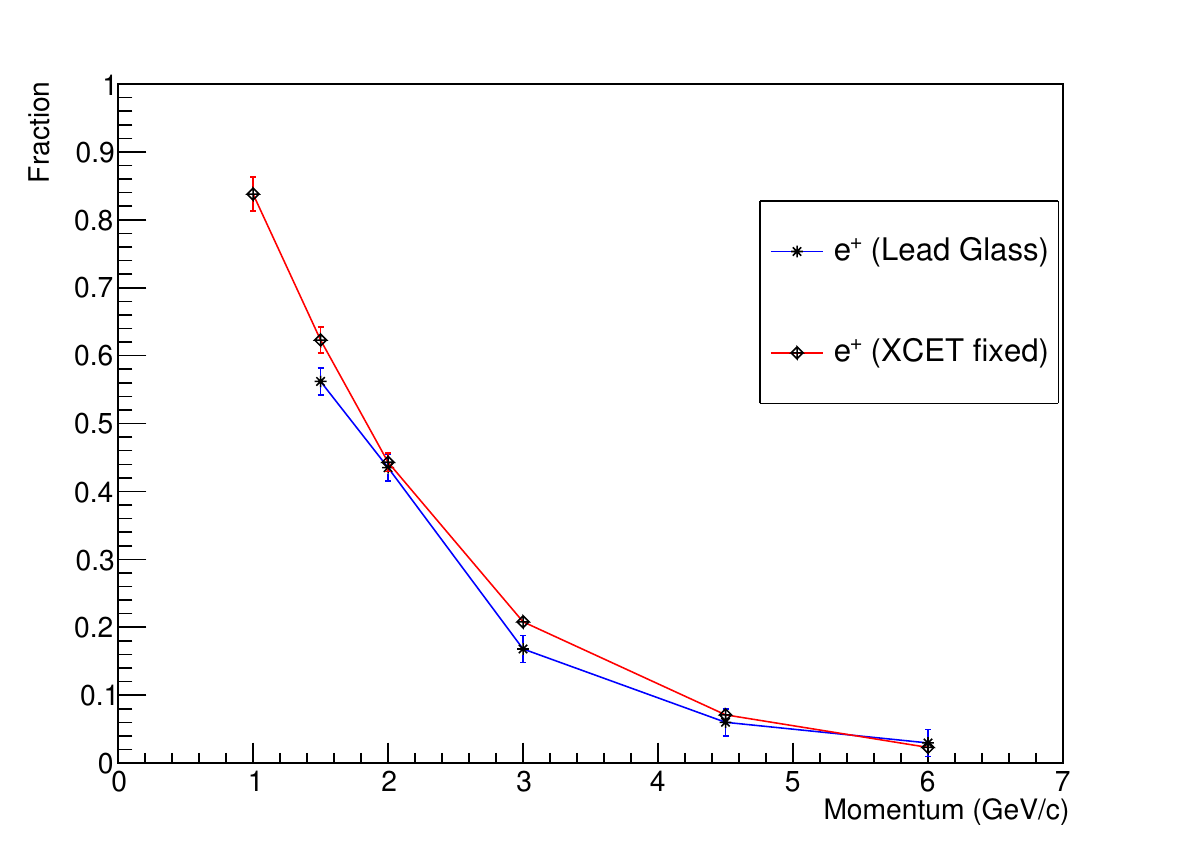}
    \includegraphics[width=0.49\textwidth]{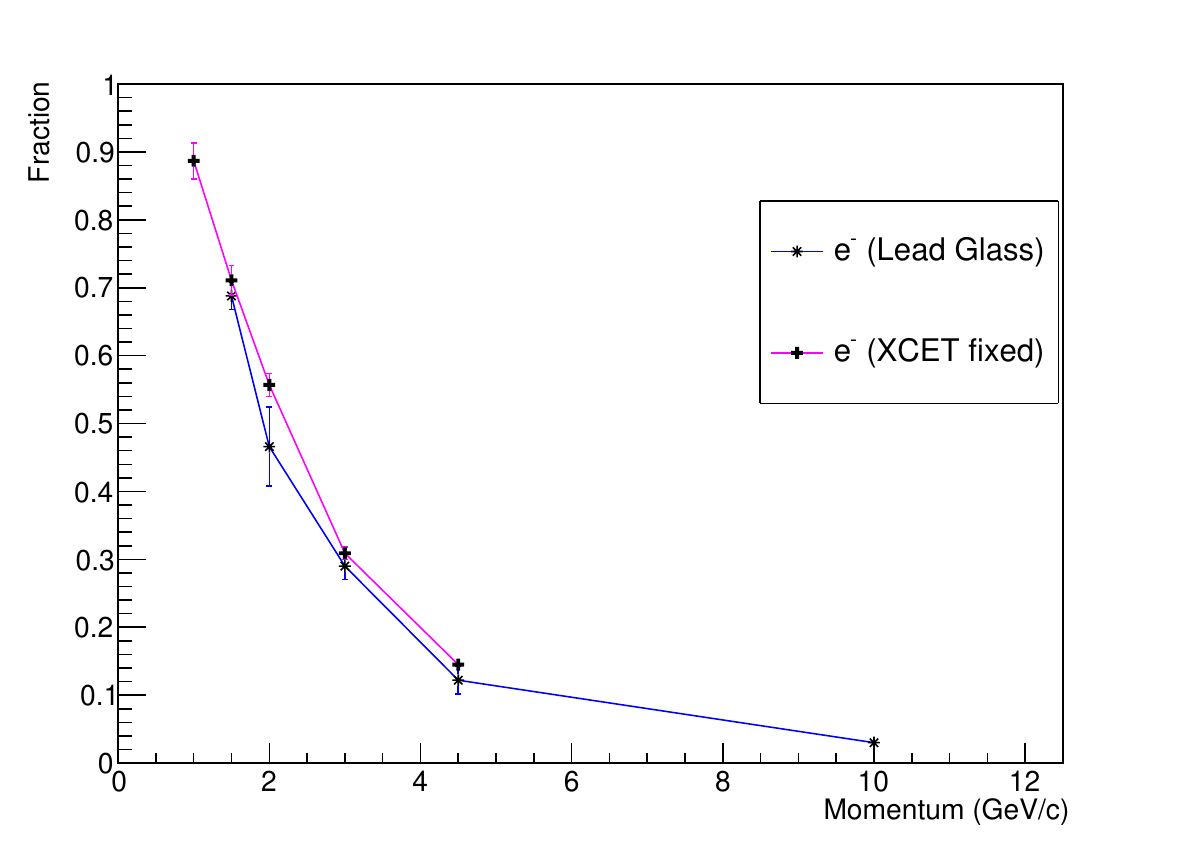}
    \caption{Comparison between the electron/positron beam fractions using the XCETs and the lead-glass detectors. Data points joined by straight lines per particle species.}
    \label{fig:L4S_LG_Comparison}
\end{figure}

Analysing the lead-glass data in runs where electrons and positrons are also identified using the XCETs, the results obtained are shown in Fig.~\ref{fig:L4S_LG_Comparison}. A very good agreement is observed, further discussed in Sec,~\ref{sec:discussion}. The results are also listed in Tab.~\ref{tab:LGresults}.

\begin{table}[htp]
\caption{Lead-Glass results}
\centering
\begin{tabular}{c|c|c}
    p (\unit{\GeVoverc})   & $e^+$               & $e^-$             \\ \hline
    1.5         & 0.562  $\pm$ 0.020  & 0.688 $\pm$ 0.020 \\
    2.0         & 0.435  $\pm$ 0.020  & 0.466 $\pm$ 0.058 \\
    3.0         & 0.168  $\pm$ 0.020  & 0.290 $\pm$ 0.020 \\
    4.5         & 0.060  $\pm$ 0.020  & 0.122 $\pm$ 0.020 \\
    6.0         & 0.030  $\pm$ 0.001  &                   \\
    10.0        &                     & 0.030 $\pm$ 0.001 \\
\end{tabular}
\label{tab:LGresults}
\end{table}

\subsection{XCET Pressure Scan Results}
\label{sec:resultspressurescan} 

\begin{figure}[!htb]
    \centering
    \includegraphics[width=0.49\textwidth]{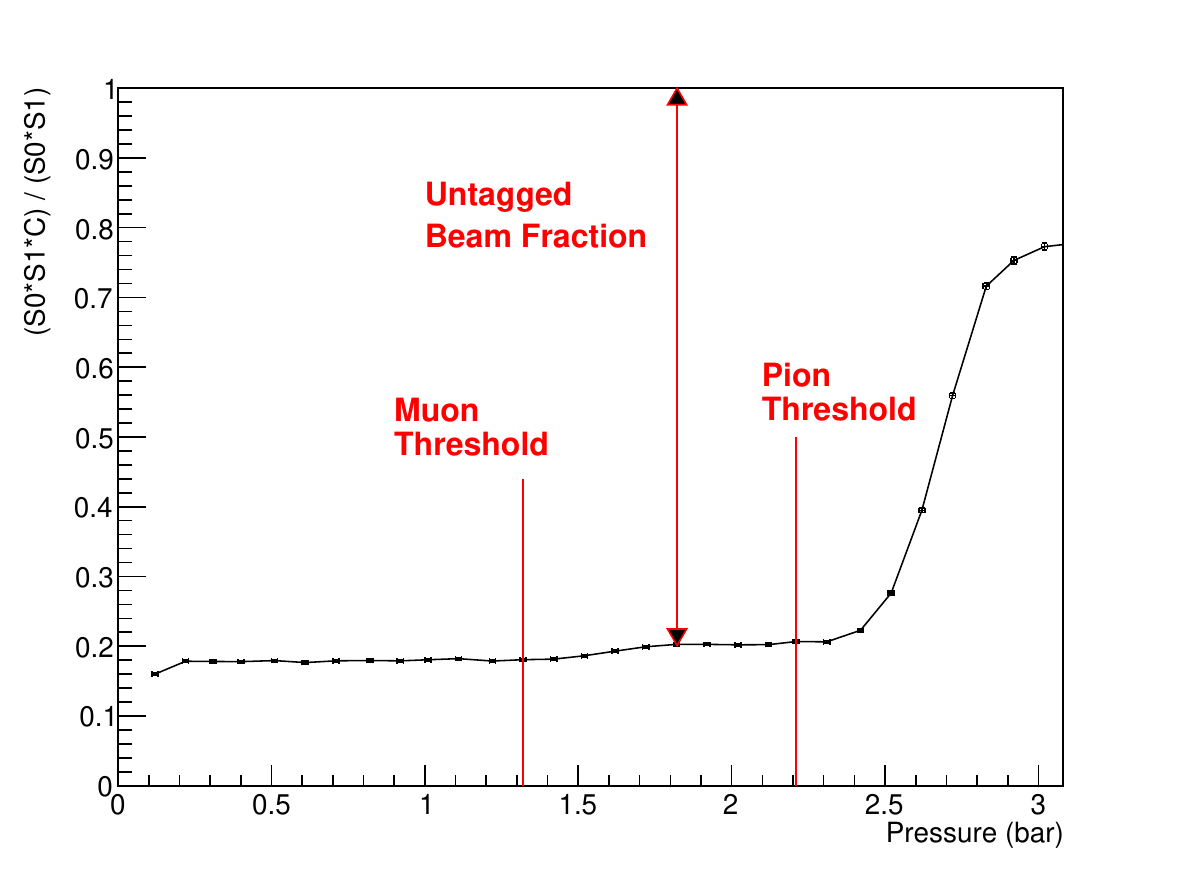}
    \includegraphics[width=0.49\textwidth]{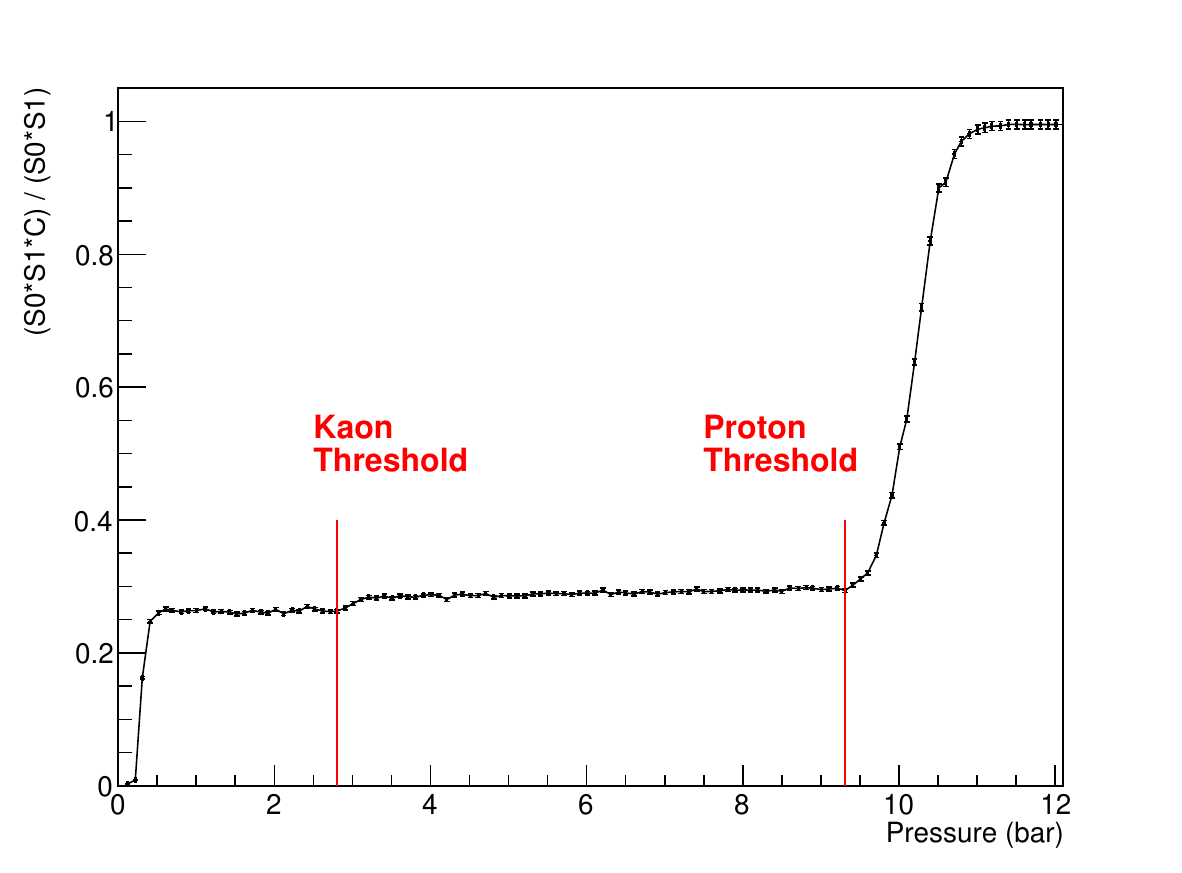}
    \caption{Pressure scan of C$_0$ at 3 (left) and \SI{10}{\GeVoverc} beam momentum (right), both for positive beam. The Y-axis is the count in the Cherenkov counter normalized to the coincidence rate of two scintillator counters in front and behind the XCET.} 
    \label{fig:pressureScan230522}
\end{figure}

Pressure scans were carried out using the high pressure XCET C$_0$ as only that device is able to reach the higher pressure allowing for proton and kaon identification over the widest range. The coincidence of scintillator counters S$_0$ and S$_1$ is used as the trigger. It is noted this is different from the previous setups, and is also discussed in Sec.~\ref{sec:trigger_comparison}. Figure~\ref{fig:pressureScan230522} shows an example of pressure scans taken at 3 and at \SI{10}{\GeVoverc}. Particle fractions were calculated from the difference in coincidence rates between the plateaus. Statistical errors were derived from the counts per pressure step, systematic errors are included as discussed below. The particle fraction is close to 1 after the proton step, indicating that the Cherenkov counter is almost fully efficient. The untagged beam fraction at \SI{3}{\GeVoverc}, marked by an arrow in the left-hand plot, measures the fraction of particles which have not yet reached their threshold; in this case pions, kaons, and protons. 
 
The two pressure scans shown in Fig.~\ref{fig:pressureScan230522} showcase a number of important features. At low pressure, the increase of pressure needed to reach full efficiency is small. The electron step at \SI{3}{\GeVoverc}, for all intents and purposes located at zero, rises to full efficiency in \SI{0.2}{\bar}. The pion step in the same plot requires about \SI{0.7}{\bar}, the proton step in the plot at \SI{10}{\GeVoverc} requires about \SI{1.7}{\bar}. This effect is mostly attributed to the chromatic dispersion effect described in Sec.~\ref{sec:XCET_refractive_index}, which scales with absolute pressure.  

\begin{figure}[!htb]
    \centering
    \includegraphics[width=0.95\textwidth]{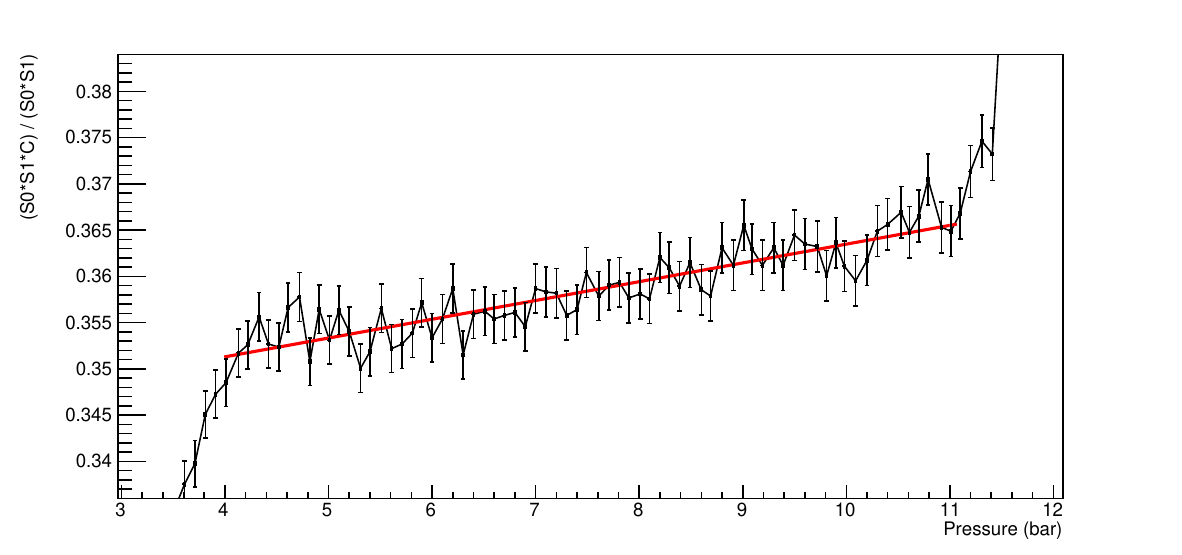}
    \caption{Detailed view of pressure scan plateau between kaon and proton thresholds at \SI{9}{\GeVoverc}, taken with high pressure XCET C$_0$.} 
    \label{fig:slopeFit9GeVpositive}
\end{figure}

A closer inspection of the pressure scans, e.g., in Fig.~\ref{fig:pressureScan230522} shows that the plateau between the kaon and proton thresholds has a positive slope. This is attributed to the creation of $\delta$-electrons by particles that have not yet crossed the Cherenkov threshold. Such electrons can generate detectable Cherenkov light even for particles that do not. This effect is expected to increase linearly with pressure and the untagged beam fraction, see Fig.~\ref{fig:pressureScan230522}. Such a region is shown in Fig.~\ref{fig:slopeFit9GeVpositive}, which features a linear fit at the plateau between pion and kaon thresholds at \SI{9}{\GeVoverc}. Other explanations are not excluded, such as detector noise or a further wide-spectrum muon contribution, but neither of these are consistent with the observation over many datasets of a linear slope as a function of pressure. 

\begin{figure}[!htb]
	\centering
	\includegraphics[width=0.49\textwidth]{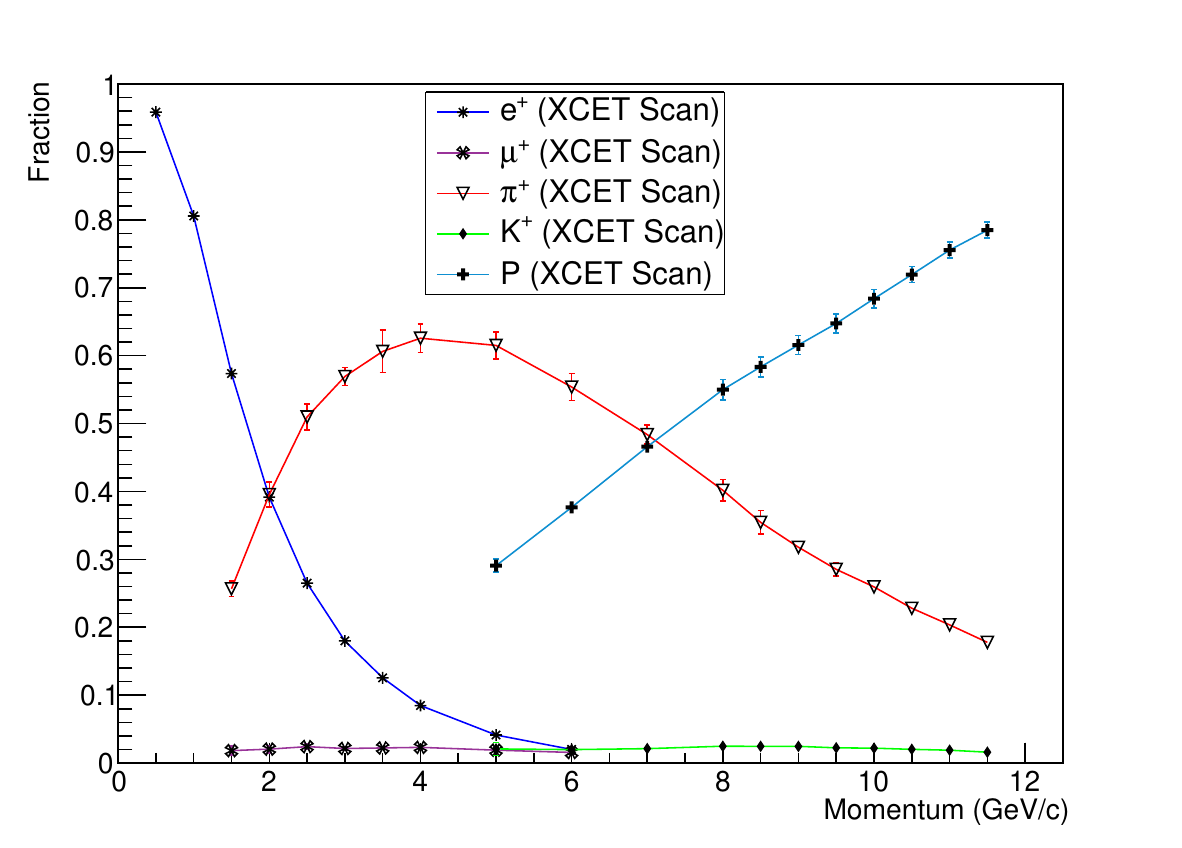}
	\includegraphics[width=0.49\textwidth]{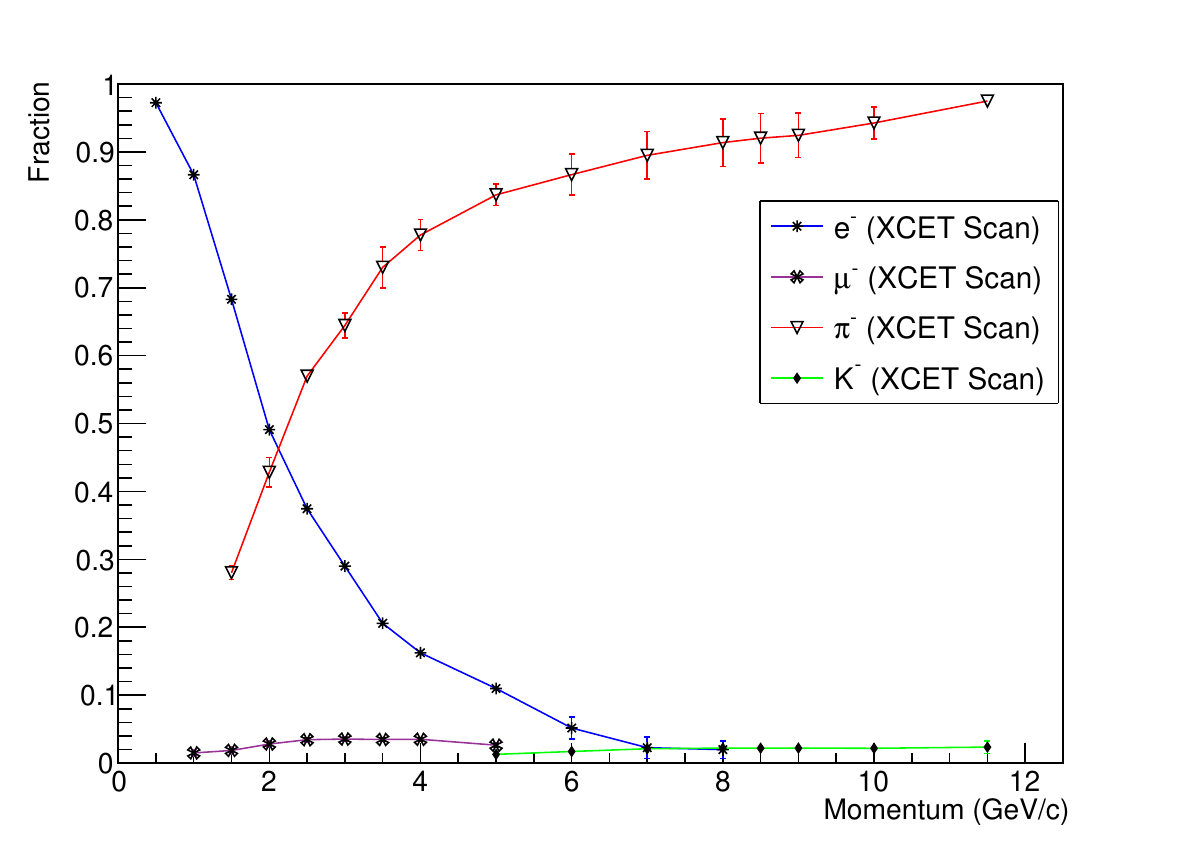}
	\caption{Particle contents for positive particles (left) and negative particles (right). Data points joined by straight lines per particle species. Anti-protons were not included in the analysis. Not all particles are covered over the full momentum range as the threshold pressures are not accessible (e.g. protons of \SI{4}{\GeVoverc} or below, or the pressure steps of the various populations merge (e.g. electrons over \SI{7}{\GeVoverc}).} 
	\label{fig:positiveParticlesPressureScan}
\end{figure}

Pressure scans from test beam periods at T10 in May 2023, August 2023, and March 2025 have been analysed. In the period between 2023 and 2025, some of the components of the high pressure XCET C$_0$ were replaced, notably the \SI{45}{\degree} mirror for one with better UV reflectivity leading to more efficient photon collection. This shifted the apparent thresholds, as the UV light becomes detectable already at lower pressures due to chromatic dispersion, as set out in Sec.~\ref{sec:XCET_refractive_index}. This is not an issue as the size of the pressure steps is independent of the pressure at which they occur, and the absolute precise threshold value does not enter the analysis. Due to the slow rise of the plateaus caused by $\delta$-electrons, the threshold steps are corrected to obtain the particle contents. This was done by performing linear fits along a plateau between threshold steps. This is not possible in all cases due to the very small pressure interval between thresholds and limited statistics. This correction is interpreted as a systematic error on the particle content measurements. From linear fits to plateaus across the dataset the average slope, normalised to the untagged beam fraction, was found to be \SI[separate-uncertainty = true]{3.20(11)e-3}{\per\bar}, per fraction of untagged beam. This number was found to be consistent across the 2023 and 2025 datasets. As an example, across a \SI{3}{\bar} plateau with a height of 0.25, this would imply an absolute systematic error of 0.0072 on the particle fraction it precedes. Statistical errors on the derived particle fractions are also computed from these linear fits as the error on the intercept. The results for the pressure scans are shown in Fig.~\ref{fig:positiveParticlesPressureScan} and Tabs.~\ref{tab:PositiveScanResults} and \ref{tab:NegativeScanResults}.

\begin{table}[htp]
\caption{XCET Scan Results, Positive particles}
\centering
\resizebox{\textwidth}{!}{
\begin{tabular}{c|c|c|c|c|c}
    p (\unit{\GeVoverc})   & $e^+$              & $\mu^+$           & $\pi^+$           & $K^+$             & P                 \\ \hline
    0.5         & 0.959 $\pm$ 0.004  &                   &                   &                   &                   \\
    1.0         & 0.806 $\pm$ 0.003  &                   &                   &                   &                   \\
    1.5         & 0.574 $\pm$ 0.001  & 0.018 $\pm$ 0.007 & 0.257 $\pm$ 0.012 &                   &                   \\
    2.0         & 0.392 $\pm$ 0.001  & 0.021 $\pm$ 0.005 & 0.395 $\pm$ 0.019 &                   &                   \\
    2.5         & 0.265 $\pm$ 0.001  & 0.024 $\pm$ 0.004 & 0.510 $\pm$ 0.019 &                   &                   \\
    3.0         & 0.180 $\pm$ 0.001  & 0.021 $\pm$ 0.003 & 0.569 $\pm$ 0.013 &                   &                   \\
    3.5         & 0.126 $\pm$ 0.001  & 0.022 $\pm$ 0.002 & 0.606 $\pm$ 0.031 &                   &                   \\
    4.0         & 0.085 $\pm$ 0.003  & 0.023 $\pm$ 0.003 & 0.626 $\pm$ 0.021 &                   &                   \\
    5.0         & 0.041 $\pm$ 0.002  & 0.019 $\pm$ 0.002 & 0.615 $\pm$ 0.020 & 0.021 $\pm$ 0.009 & 0.291 $\pm$ 0.010 \\
    6.0         & 0.020 $\pm$ 0.002  & 0.016 $\pm$ 0.002 & 0.554 $\pm$ 0.020 & 0.020 $\pm$ 0.008 & 0.377 $\pm$ 0.008 \\
    7.0         &                    &                   & 0.484 $\pm$ 0.014 & 0.021 $\pm$ 0.007 & 0.466 $\pm$ 0.007 \\
    8.0         &                    &                   & 0.402 $\pm$ 0.016 & 0.025 $\pm$ 0.006 & 0.550 $\pm$ 0.015 \\
    8.5         &                    &                   & 0.355 $\pm$ 0.017 & 0.025 $\pm$ 0.006 & 0.584 $\pm$ 0.015 \\
    9.0         &                    &                   & 0.318 $\pm$ 0.008 & 0.025 $\pm$ 0.006 & 0.616 $\pm$ 0.014 \\
    9.5         &                    &                   & 0.285 $\pm$ 0.009 & 0.022 $\pm$ 0.006 & 0.647 $\pm$ 0.014 \\
    10.0        &                    &                   & 0.259 $\pm$ 0.007 & 0.022 $\pm$ 0.006 & 0.684 $\pm$ 0.014 \\
    10.5        &                    &                   & 0.228 $\pm$ 0.007 & 0.020 $\pm$ 0.005 & 0.719 $\pm$ 0.012 \\
    11.0        &                    &                   & 0.204 $\pm$ 0.002 & 0.019 $\pm$ 0.005 & 0.756 $\pm$ 0.012 \\
    11.5        &                    &                   & 0.178 $\pm$ 0.002 & 0.016 $\pm$ 0.004 & 0.785 $\pm$ 0.012 \\
\end{tabular}
}
\label{tab:PositiveScanResults}
\end{table}

\begin{table}[htp]
\caption{XCET Scan Results, Negative particles}
\centering
\resizebox{\textwidth}{!}{
\begin{tabular}{c|c|c|c|c}
    p (\unit{\GeVoverc})	& $e^-$             & $\mu^-$           & $\pi^-$	        & $K^-$             \\  
    \hline
    0.5	        & 0.973 $\pm$ 0.001 &                   &	                &                   \\
    1.0	        & 0.867 $\pm$ 0.001 & 0.015 $\pm$ 0.005 &	                &                   \\
    1.5	        & 0.683 $\pm$ 0.001 & 0.019 $\pm$ 0.005 & 0.280 $\pm$ 0.010 &                   \\
    2.0	        & 0.491 $\pm$ 0.001 & 0.028 $\pm$ 0.005 & 0.428 $\pm$ 0.022 &                   \\
    2.5	        & 0.375 $\pm$ 0.001 & 0.034 $\pm$ 0.003 & 0.570 $\pm$ 0.008 &                   \\
    3.0	        & 0.290 $\pm$ 0.002 & 0.035 $\pm$ 0.003 & 0.645 $\pm$ 0.018 &                   \\
    3.5	        & 0.206 $\pm$ 0.001 & 0.035 $\pm$ 0.002 & 0.730 $\pm$ 0.030 &                   \\
    4.0	        & 0.162 $\pm$ 0.002 & 0.035 $\pm$ 0.002 & 0.778 $\pm$ 0.023 &                   \\
    5.0	        & 0.110 $\pm$ 0.005 & 0.026 $\pm$ 0.003 & 0.837 $\pm$ 0.016 & 0.013 $\pm$ 0.002 \\
    6.0	        & 0.052 $\pm$ 0.016 &                   & 0.867 $\pm$ 0.030 & 0.017 $\pm$ 0.002 \\
    7.0	        & 0.023 $\pm$ 0.016 &                   & 0.895 $\pm$ 0.035 & 0.021 $\pm$ 0.002 \\
    8.0	        & 0.020 $\pm$ 0.013 &                   & 0.914 $\pm$ 0.035 & 0.022 $\pm$ 0.002 \\
    8.5	        &                   &                   & 0.920 $\pm$ 0.036 & 0.022 $\pm$ 0.004 \\
    9.0	        &	                &                   & 0.924 $\pm$ 0.033 & 0.022 $\pm$ 0.004 \\
    10.0        &	                &                   & 0.943 $\pm$ 0.024 & 0.022 $\pm$ 0.006 \\
    11.5	    &	                &                   & 0.975 $\pm$ 0.007 & 0.023 $\pm$ 0.009 \\
\end{tabular}
}
\label{tab:NegativeScanResults}
\end{table}

\subsection{Trigger Comparison}
\label{sec:trigger_comparison}
The trigger definition for the various measurements described below can be divided into two parts. For the fixed XCET measurements and the lead-glass calorimeter measurements, the hardware trigger was defined as the coincidence of S$_1$ and S$_4$. For the XCET pressure scans, the trigger was the coincidence of S$_0$ and S$_1$. The scintillator counter S$_4$ is placed around \SI{8}{\meter} downstream (see Fig.~\ref{fig:pakistanExperimentalLayout}). This leads to scattering and interactions, and consequently a difference in the beam composition, in particular affecting the electron respectively positron component of the beam interacting with the air. The dominant interaction leading to energy loss for high-energy electrons is Bremsstrahlung, considering here only \SI{0.5}{\GeVoverc} and higher. Coulomb scattering is not expected to be relevant due to the large size of the downstream detectors. Counter S$_4$ measures 30$\times$\SI{20}{\centi\meter\squared} and the lead-glass calorimeter behind it covers a larger area still. The analysis for the lead-glass calorimeter uses the sum of the signal on all calorimeter crystals. As the angles under which the Bremsstrahlung photons are emitted are very small, the sum of the nine crystals will likely also catch these, and the summation will fall in the electron peak still.

The amount of air between S$_1$ and S$_4$ corresponds to a percent-level fraction of a nuclear interaction length, relevant for hadrons. Still, almost any interaction produced by this will trigger the large scintillator counter S$_4$. It is thus expected that there is either no or only a very small effect of the difference in trigger definition of the two setups. No substantial effect on the measured beam composition is expected, even though the material in between is not negligible. 

For the ToF measurements, the trigger was the coincidence of S$_1$ and S$_4$. The ToF difference was calculated from S$_0$ to S$_4$. This choice effectively means the trigger consists of the coincidence of S$_0$, S$_1$ and S$_4$ The difference with the trigger only consisting of S$_1$ and S$_4$ was investigated by checking the electron content also at these settings, for the two different trigger definitions, as this information is readily available from the calorimeter. With this redefined trigger, \SI{0.8}{\percent} more electrons are found at \SI{1}{\GeVoverc} and \SI{0.6}{\percent} at \SI{2}{\GeVoverc}. These numbers are added as systematic errors on the ToF measurements as they are considered as the maximal shift possible due to the redefined trigger for the protons.

\subsection{Comparison of all achieved results}
\label{sec:compareAll} 

\begin{figure}[!htb]
	\centering
	\includegraphics[width=0.49\textwidth]{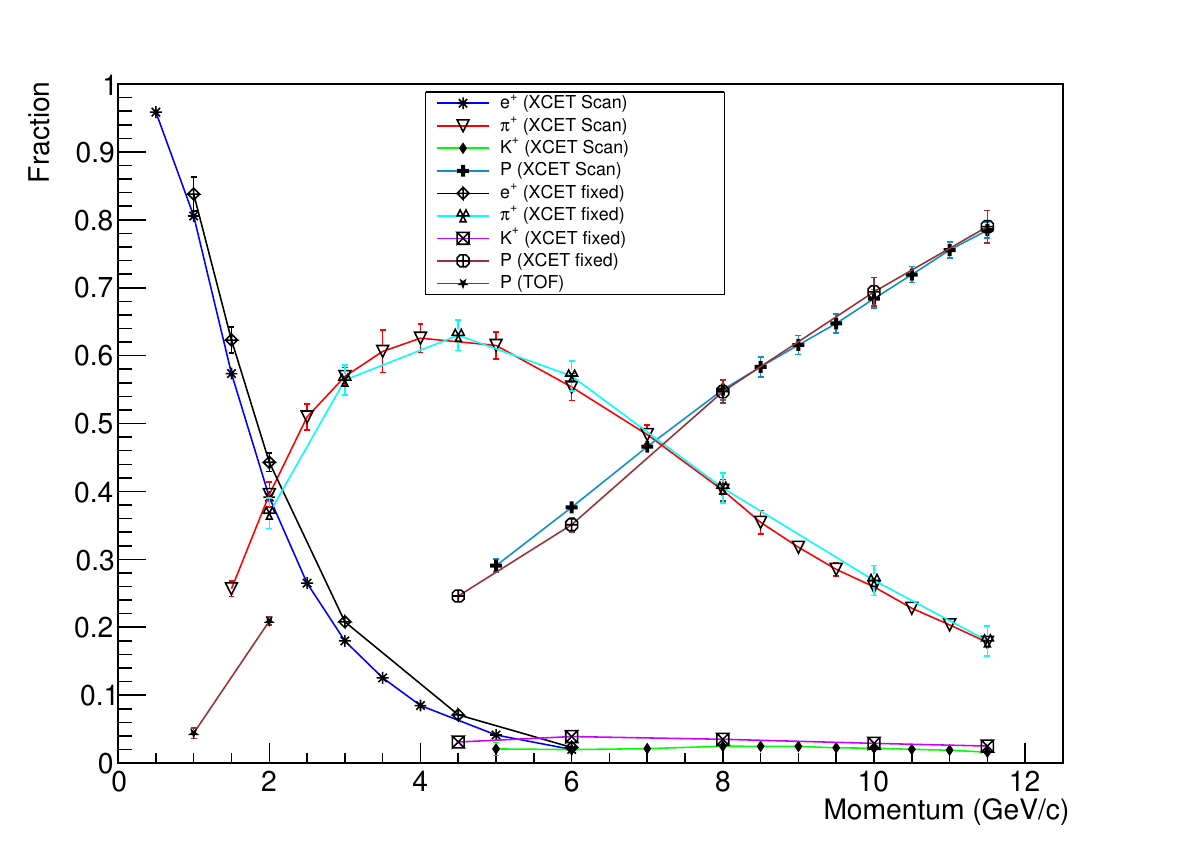}
	\includegraphics[width=0.49\textwidth]{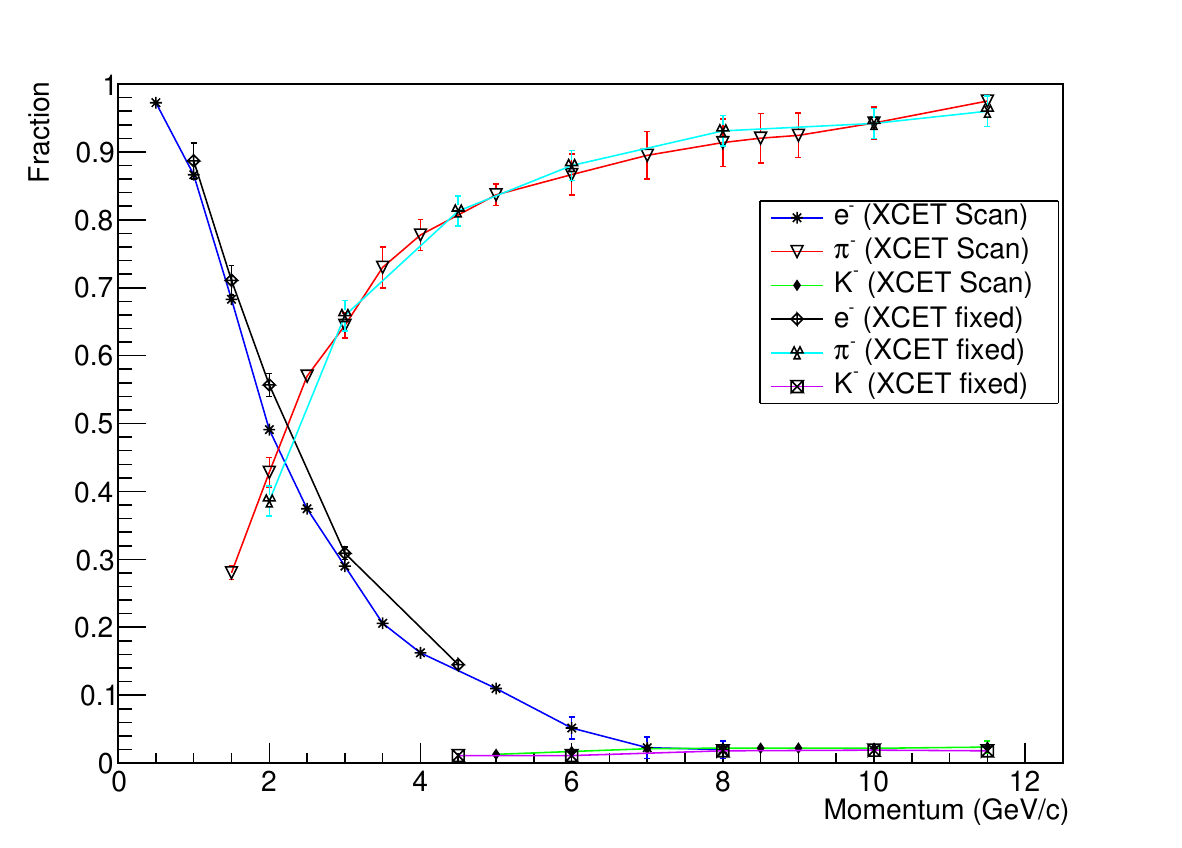}
	\caption{Full dataset for positive (left) and negative particles (right). Data points joined by straight lines per particle species.} 
	\label{fig:BL4S_Scan_Comparison_Full}
\end{figure}

The measurements from fixed-point XCET, ToF, and XCET pressure scans are compared in Fig.~\ref{fig:BL4S_Scan_Comparison_Full}. Data from the lead-glass calorimeter is omitted here as it was already compared to the fixed-point XCET in Sec.~\ref{sec:resultsLG}. 

The data for the fixed-point XCET and and for the XCET scans is in reasonable agreement. In the case of positive beam, the pion and proton distributions agree well. The positrons are found to be slightly more abundant in the fixed-point XCET measurements, ascribed to a collimator effect discussed in more detail in the following section. Similarly, the kaons appear slightly more abundant in the fixed-point XCET measurements but not to the point of substantial disagreement. In the case of negative beam, there is in general a good agreement. Only for electrons at \SI{2}{\GeVoverc} there is a substantial disagreement, as was also noted for the lead-glass calorimeter results. Broadly, the electrons are also more abundant in the fixed-point than in the XCET scan, which again is ascribed to collimator settings. It is noted that a far more complete assessment of the errors was made for the XCET scans than for the fixed-point data, and as such the XCET scan data is of higher quality even where it has larger error bars. 
\section{Discussion}
\label{sec:discussion}

\begin{figure}[!htb]
	\centering
	\includegraphics[width=0.49\textwidth]{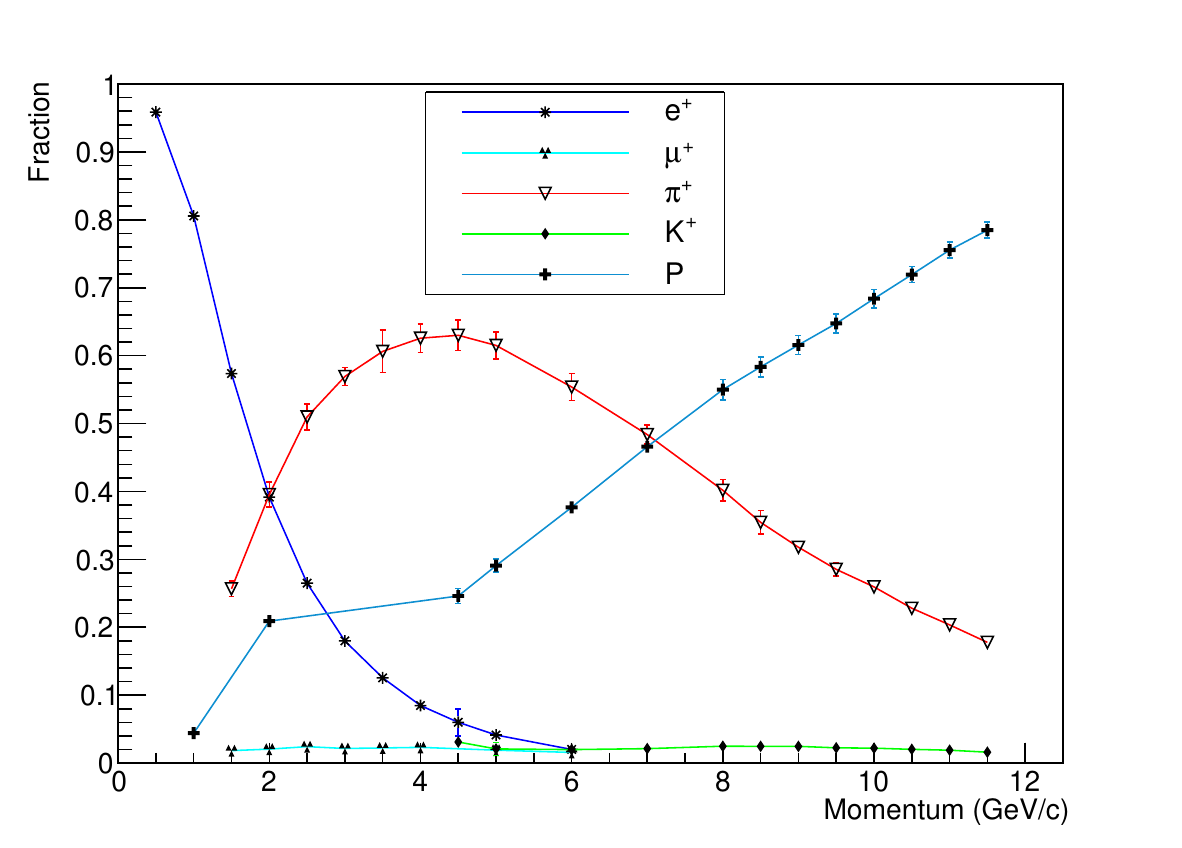}
	\includegraphics[width=0.49\textwidth]{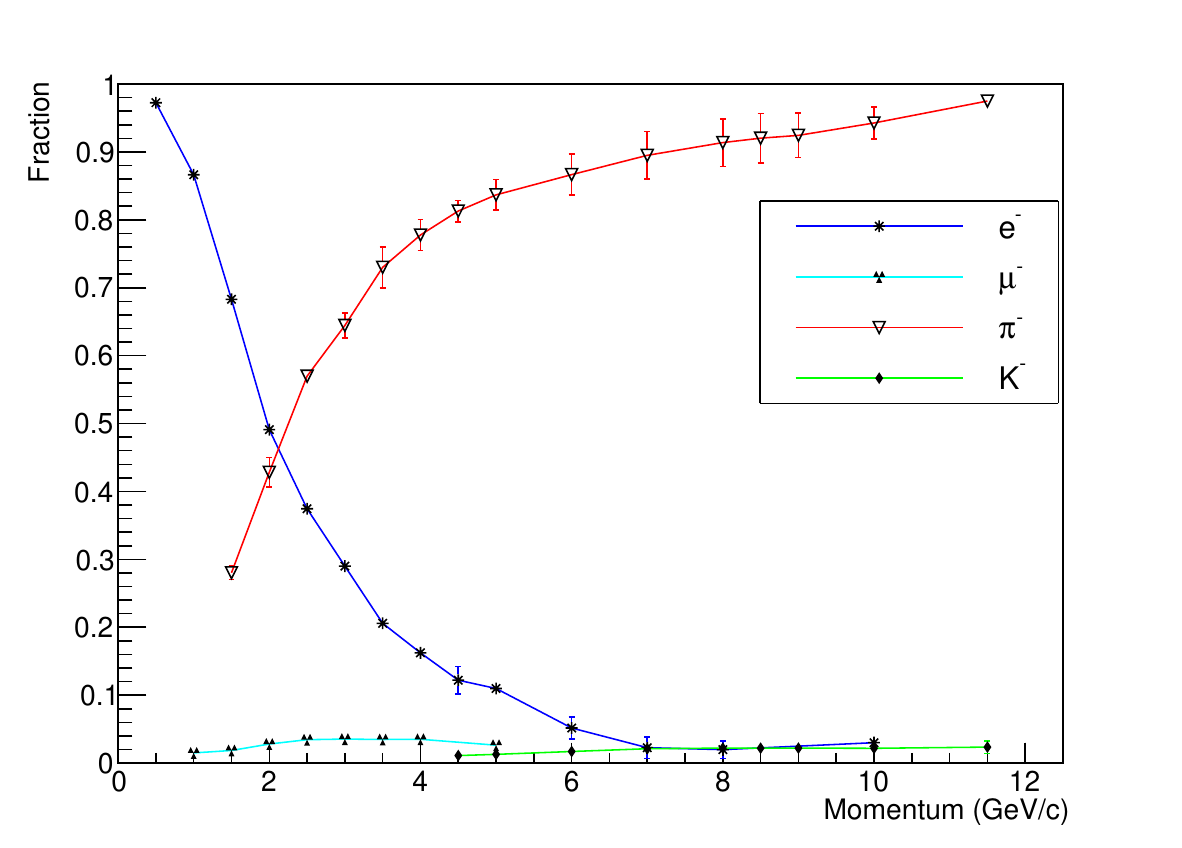}
	\caption{Particle contents for positive particles (left) and negative particles (right). Data points joined by straight lines per particle species.} 
	\label{fig:FinalBeamComposition}
\end{figure}

The fixed XCET method obviously has issues similar to the XCET pressure scan method --- there is likely a similarly increasing slope on the plateaus between the particle species turning on. Due to the method employed however, this plateau is neither observed nor can be corrected for. It would be possible to take the correction from the XCET scan method, but this approach was discarded as there is a fully separate DAQ and analysis chain. The points from the various methods are combined to derive a final set of numbers. For cases where the data from the fixed XCET method and the XCET pressure scan are present, the data from the XCET scan is listed. Where there is data present only from the lead-glass calorimeter and the fixed XCET method, the lead-glass data is preferred. The particle contents are finalized in Fig.~\ref{fig:FinalBeamComposition} and Tabs.~\ref{tab:PositiveFinalBeamComposition} and \ref{tab:NegativeFinalBeamComposition}. 

\begin{table}[htp]
\caption{Measured beam composition, positive particles. Data denoted with * from lead-glass calorimeter, ** from ToF, and *** from fixed-point XCET. Remaining points from XCET scan. }
\centering
\resizebox{\textwidth}{!}{
\begin{tabular}{c|c|c|c|c|c}
 p (\unit{\GeVoverc})   & $e^+$              & $\mu^+$           & $\pi^+$                & $K^+$               & P                    \\ \hline
    0.5         & 0.959 $\pm$ 0.004  &                   &                     &                     &                      \\
    1.0         & 0.806 $\pm$ 0.003  &                   &                     &                     & 0.044 $\pm$ 0.008**  \\
    1.5         & 0.574 $\pm$ 0.001  & 0.018 $\pm$ 0.007 & 0.257 $\pm$ 0.012   &                     &                      \\
    2.0         & 0.392 $\pm$ 0.001  & 0.021 $\pm$ 0.005 & 0.395 $\pm$ 0.019   &                     & 0.209 $\pm$ 0.006**  \\
    2.5         & 0.265 $\pm$ 0.001  & 0.024 $\pm$ 0.004 & 0.510 $\pm$ 0.019   &                     &                      \\
    3.0         & 0.180 $\pm$ 0.001  & 0.021 $\pm$ 0.003 & 0.569 $\pm$ 0.013   &                     &                      \\
    3.5         & 0.126 $\pm$ 0.001  & 0.022 $\pm$ 0.002 & 0.606 $\pm$ 0.031   &                     &                      \\
    4.0         & 0.085 $\pm$ 0.003  & 0.023 $\pm$ 0.003 & 0.626 $\pm$ 0.021   &                     &                      \\
    4.5         & 0.060 $\pm$ 0.020* &                   & 0.630 $\pm$ 0.022***& 0.031 $\pm$ 0.004***& 0.246 $\pm$ 0.008*** \\
    5.0         & 0.041 $\pm$ 0.002  & 0.019 $\pm$ 0.002 & 0.615 $\pm$ 0.020   & 0.021 $\pm$ 0.009   & 0.291 $\pm$ 0.010    \\
    6.0         & 0.020 $\pm$ 0.002  & 0.016 $\pm$ 0.002 & 0.554 $\pm$ 0.020   & 0.020 $\pm$ 0.008   & 0.377 $\pm$ 0.008    \\
    7.0         &                    &                   & 0.484 $\pm$ 0.014   & 0.021 $\pm$ 0.007   & 0.466 $\pm$ 0.007    \\
    8.0         &                    &                   & 0.402 $\pm$ 0.016   & 0.025 $\pm$ 0.006   & 0.550 $\pm$ 0.015    \\
    8.5         &                    &                   & 0.355 $\pm$ 0.017   & 0.025 $\pm$ 0.006   & 0.584 $\pm$ 0.015    \\
    9.0         &                    &                   & 0.318 $\pm$ 0.008   & 0.025 $\pm$ 0.006   & 0.616 $\pm$ 0.014    \\
    9.5         &                    &                   & 0.285 $\pm$ 0.009   & 0.022 $\pm$ 0.006   & 0.647 $\pm$ 0.014    \\
    10.0        &                    &                   & 0.259 $\pm$ 0.007   & 0.022 $\pm$ 0.006   & 0.684 $\pm$ 0.014    \\
    10.5        &                    &                   & 0.228 $\pm$ 0.007   & 0.020 $\pm$ 0.005   & 0.719 $\pm$ 0.012    \\
    11.0        &                    &                   & 0.204 $\pm$ 0.002   & 0.019 $\pm$ 0.005   & 0.756 $\pm$ 0.012    \\
    11.5        &                    &                   & 0.178 $\pm$ 0.002   & 0.016 $\pm$ 0.004   & 0.785 $\pm$ 0.012    \\
\end{tabular}
}
\label{tab:PositiveFinalBeamComposition}
\end{table}

\begin{table}[htp]
\caption{Measured beam composition, negative particles. Data denoted with * from lead-glass calorimeter and *** from fixed-point XCET. Remaining points from XCET scan.}
\centering
\resizebox{\textwidth}{!}{
\begin{tabular}{c|c|c|c|c}
     p (\unit{\GeVoverc})	& $e^-$             & $\mu^-$           & $\pi^-$	         & $K^-$               \\  
    \hline
    0.5	        & 0.973 $\pm$ 0.001 &                   &	                  &                     \\
    1.0	        & 0.867 $\pm$ 0.001 & 0.015 $\pm$ 0.005 &	                  &                     \\
    1.5	        & 0.683 $\pm$ 0.001 & 0.019 $\pm$ 0.005 & 0.280 $\pm$ 0.010   &                     \\
    2.0	        & 0.491 $\pm$ 0.001 & 0.028 $\pm$ 0.005 & 0.428 $\pm$ 0.022   &                     \\
    2.5	        & 0.375 $\pm$ 0.001 & 0.034 $\pm$ 0.003 & 0.570 $\pm$ 0.008   &                     \\
    3.0	        & 0.290 $\pm$ 0.002 & 0.035 $\pm$ 0.003 & 0.645 $\pm$ 0.018   &                     \\
    3.5	        & 0.206 $\pm$ 0.001 & 0.035 $\pm$ 0.002 & 0.730 $\pm$ 0.030   &                     \\
    4.0	        & 0.162 $\pm$ 0.002 & 0.035 $\pm$ 0.002 & 0.778 $\pm$ 0.023   &                     \\
    4.5         & 0.122 $\pm$ 0.020*&                   & 0.813 $\pm$ 0.022***& 0.011 $\pm$ 0.005***\\
    5.0	        & 0.110 $\pm$ 0.005 & 0.026 $\pm$ 0.003 & 0.837 $\pm$ 0.016   & 0.013 $\pm$ 0.002   \\
    6.0	        & 0.052 $\pm$ 0.016 &                   & 0.867 $\pm$ 0.030   & 0.017 $\pm$ 0.002   \\
    7.0	        & 0.023 $\pm$ 0.016 &                   & 0.895 $\pm$ 0.035   & 0.021 $\pm$ 0.002   \\
    8.0	        & 0.020 $\pm$ 0.013 &                   & 0.914 $\pm$ 0.035   & 0.022 $\pm$ 0.002   \\
    8.5	        &                   &                   & 0.920 $\pm$ 0.036   & 0.022 $\pm$ 0.004   \\
    9.0	        &	                &                   & 0.924 $\pm$ 0.033   & 0.022 $\pm$ 0.004   \\
    10.0        & 0.03 $\pm$ 0.001* &                   & 0.943 $\pm$ 0.024   & 0.022 $\pm$ 0.006   \\
    11.5	    &	                &                   & 0.975 $\pm$ 0.006   & 0.023 $\pm$ 0.009   \\
\end{tabular}
}
\label{tab:NegativeFinalBeamComposition}
\end{table}

Figure~\ref{fig:FinalBeamComposition} and Tabs.~\ref{tab:PositiveFinalBeamComposition} and \ref{tab:NegativeFinalBeamComposition} give an overview of the particle content in the T10 beam. Where the same particle species is assessed through multiple methods, they are found to be consistent within error bars. A substantial part of the particle and momentum space remains uncovered. As momentum increases, the pressure steps of the lighter particle species merge and can no longer be measured with an XCET. The electron respectively positron spectrum at higher energies could be covered by a calorimeter. Likely, the best method to derive the kaon and proton content at low momenta would be ToF. These are left as possibilities for a future experiment. 

The full measurement of the muon component of the beam is more difficult. Muons can be deflected in magnet yokes, and as such have a broader range of paths they could follow to reach the experimental area. Similarly, pions decaying along the length of the beam line will add a further, wide spectrum. The muon component as included in the analysis does not cover the full momentum range. For the muons, only data from the XCET scans was included, and data sets where no clear step in the pressure scan could be distinguished were omitted for this particle species. Within the present study, no clear path was identified that would allow for better muon identification. 

Some particle species were not considered in this analysis. It is known that there is an anti-proton component in the negative beam, expected to be at a sub-percent level for most momenta considered. The fixed XCET method would be suitable for this at high momenta. At low momenta, the ToF method could offer further insight. Nevertheless, this determination was not considered as a key target for this analysis. Similarly, in the ToF analysis at low energy a sub-percent fraction of deuterons was identified in the beam but not investigated any further. 

The slope found on the plateaus of the XCET pressure scans was not fully resolved. The proposed hypothesis could be tested to a reasonable degree in simulation. Still, as its addition to the error analysis led to sub-percent errors in most cases, it was not deemed to be needed in the present study. Additionally, in some of the pressure scans where a very long plateau was present, it was noted that a slight negative gradient was visible, unexpectedly. No clear attribution can be made for this. In the cases where this effect was present, it was found to be small and in line with already assessed errors. No further additional error was added. 

During the final data collection in 2025 it was observed that the collimator slits affect the beam composition. An effect of this kind was anticipated for the collimator defining the momentum acceptance, and this opening was maintained at $\pm$\SI{1}{\percent} through all data taking. The upstream acceptance collimators were varied. For the fixed-point XCET data taking, the rates were typically some $10^4$, for the XCET pressure scans usually 1--\num{2e5}. In later tests where the acceptance collimator was varied, a shift of a few percent in the electron content was noted depending on the opening of the acceptance collimators. It implies not all production of electrons in particular comes from the primary target when the acceptance collimators are more closed. A full investigation into this effect was deemed beyond the scope of this work, but it is noted for future experimentation. 

\section{Conclusion and Future Work}

The present work has established the particle content of the T10 secondary beam line of the CERN PS East Area derived from the beryllium/tungsten primary target, with which the line operates most of the time.

Future work will focus on the simulation of the full beam line and to investigate the relative particle content in more detail, and to cover the full parameter space. As opportunity arises, the current knowledge will be supplemented with not-yet established particle species, such as the anti-proton content of the negative beam. Collaboration with future testbeam users will be sought out to fill the remaining parameter space.
\section*{Acknowledgments }

Beamline for Schools (BL4S) is a physics competition for high school students from all around the world organized at CERN and DESY, Germany \cite{bl4s_homepage}. One of the winning teams in BL4S 2023, team ``Particular Perspective'' proposed to measure the particle composition of the T10 beam line. This paper, partially built on data collected in 2023 with the student team, is the result of their proposal. The authors extend their heartfelt thanks to the members of Team Particular Perspective for their contributions: M.~Akhtar (coach), M.S.~Tarar, W.~Ahmed, M.S.~Bilal, M.M.A.~Maitla, A.W.~Akram, M.H.~Haider, M.A.~Abdali, M.Z.~Abbas and M.A.~Masood. The authors express their deep gratitude for the extensive efforts of M.~Joos as BL4S technical coordinator. The authors kindly thank N.~Charitonidis and L.~Gatignon for invaluable discussion and comments during the measurements, analysis and paper preparation. The authors thank the Beam Instrumentation team for the careful preparation and maintenance of the beam line detectors. The authors thank the East Area superintendent, A.~Ebn~Rahmoun, and all CERN technical teams, for the help provided. BL4S is funded by the CERN \& Society Foundation. The Beamline for Schools (BL4S) competition is made possible thanks to the generous donations received from Rolex, the Wilhelm and Else~Heraeus Foundation and other supporters of the CERN \& Society Foundation. The authors express their gratitude to CERN for the use of the facilities and resources.

\bibliographystyle{style/my-lim-num}
\bibliography{bib/reportpressureScanBL4S}

\begin{thebibliography}{10}
\expandafter\ifx\csname url\endcsname\relax
  \def\url#1{\texttt{#1}}\fi
\expandafter\ifx\csname urlprefix\endcsname\relax\def\urlprefix{URL }\fi
\expandafter\ifx\csname href\endcsname\relax
  \def\href#1#2{#2} \def\path#1{#1}\fi

\bibitem{Bernhard:2792490}
J.~Bernhard, P.~Burdelski, B.~Carlsen, et~al., {CERN Proton Synchrotron East Area Facility: Upgrades and renovation during Long Shutdown 2}, Tech. rep., CERN, Geneva (2021).
\newblock \href {http://dx.doi.org/10.23731/CYRM-2021-004} {\path{doi:10.23731/CYRM-2021-004}}.

\bibitem{Gatignon:2730780}
L.~Gatignon, {Design and Tuning of Secondary Beamlines in the CERN North and East Areas}, {CERN Document Server}\href {http://dx.doi.org/10.17181/CERN.T6FT.6UDG} {\path{doi:10.17181/CERN.T6FT.6UDG}}.

\bibitem{bl4s_homepage}
{BL4S Homepage, \href{https://beamlineforschools.cern/}{https://beamlineforschools.cern/}, last accessed: 27/06/2025}.

\bibitem{rae:icalepcs2021-tupv047}
B.~Rae, V.~Baggiolini, D.~Banerjee, et~al., {Controlling the CERN Experimental Area Beams}, in: Proc. ICALEPCS'21, no.~18 in International Conference on Accelerator and Large Experimental Physics Control Systems, JACoW Publishing, Geneva, Switzerland, 2022, pp. 509--513.
\newblock \href {http://dx.doi.org/10.18429/JACoW-ICALEPCS2021-TUPV047} {\path{doi:10.18429/JACoW-ICALEPCS2021-TUPV047}}.

\bibitem{BRUN199781}
R.~Brun, F.~Rademakers, {ROOT — An object oriented data analysis framework}, Nuclear Instruments and Methods in Physics Research Section A: Accelerators, Spectrometers, Detectors and Associated Equipment 389~(1) (1997) 81--86, new Computing Techniques in Physics Research V.
\newblock \href {http://dx.doi.org/10.1016/S0168-9002(97)00048-X} {\path{doi:10.1016/S0168-9002(97)00048-X}}.

\bibitem{CHARITONIDIS201920}
N.~Charitonidis, P.~Chatzidaki, Y.~Karyotakis, M.~Rosenthal, {Particle identification in the low-GeV/c regime using Octafluoropropane (R-218) as Cherenkov radiator}, Nuclear Instruments and Methods in Physics Research Section B: Beam Interactions with Materials and Atoms 457 (2019) 20--23.
\newblock \href {http://dx.doi.org/10.1016/j.nimb.2019.07.014} {\path{doi:10.1016/j.nimb.2019.07.014}}.

\bibitem{CHARITONIDIS2017134}
N.~Charitonidis, Y.~Karyotakis, L.~Gatignon, \href{https://www.sciencedirect.com/science/article/pii/S0168583X17308133}{{Estimation of the R134a gas refractive index for use as a Cherenkov radiator, using a high energy charged particle beam}}, Nuclear Instruments and Methods in Physics Research Section B: Beam Interactions with Materials and Atoms 410 (2017) 134--138.
\newblock \href {http://dx.doi.org/https://doi.org/10.1016/j.nimb.2017.08.020} {\path{doi:https://doi.org/10.1016/j.nimb.2017.08.020}}.
\newline\urlprefix\url{https://www.sciencedirect.com/science/article/pii/S0168583X17308133}

\bibitem{BIDEAUMEHU1973432}
A.~Bideau-Mehu, Y.~Guern, R.~Abjean, A.~Johannin-Gilles, Interferometric determination of the refractive index of carbon dioxide in the ultraviolet region, Optics Communications 9~(4) (1973) 432--434.
\newblock \href {http://dx.doi.org/10.1016/0030-4018(73)90289-7} {\path{doi:10.1016/0030-4018(73)90289-7}}.

\bibitem{buesa_orgaz:ipac2024-wepg90}
J.~Buesa~Orgaz, et~al., {Reflectivity studies and production of new flat mirrors for the Cherenkov threshold detectors at CERN}, in: Proc. IPAC'24, no.~15 in IPAC'24 - 15th International Particle Accelerator Conference, JACoW Publishing, Geneva, Switzerland, 2024, pp. 2434--2437.
\newblock \href {http://dx.doi.org/10.18429/JACoW-IPAC2024-WEPG90} {\path{doi:10.18429/JACoW-IPAC2024-WEPG90}}.

\bibitem{bib:bl4sBeamComposition}
A.~Hayat, B.~Gokturk, M.~Schwinzerl, J.~Petersen, {{Beamline for Schools 2023 Project Technical Report: Beam Composition T10}}, Tech. Rep. \href{https://edms.cern.ch/document/3304505}{EDMS 3304505}, CERN, Geneva (2025).

\bibitem{XCET_electronics}
A.~Manarin, {{Module VME XCET Ou eXperimental CErenkov Threshold counter -- Manuel d'Information Technique (Révision 1.2-XCET EDA-00032-V3)}}, Tech. Rep. \href{https://edms.cern.ch/document/2806271/}{EDMS 2806271}, CERN, Geneva (2012).

\bibitem{1991275}
{The OPAL collaboration}, {The OPAL detector at LEP}, Nuclear Instruments and Methods in Physics Research Section A: Accelerators, Spectrometers, Detectors and Associated Equipment 305~(2) (1991) 275--319.
\newblock \href {http://dx.doi.org/10.1016/0168-9002(91)90547-4} {\path{doi:10.1016/0168-9002(91)90547-4}}.

\end{thebibliography}

\end{document}